\documentclass[fleqn,10pt]{wlscirep}
\usepackage[utf8]{inputenc}
\usepackage[T1]{fontenc}
\usepackage{xcolor}
\usepackage{caption}
\usepackage{subcaption}
\usepackage{graphicx}
\usepackage[normalem]{ulem}

\usepackage{xr}
\makeatletter
\newcommand*{\addFileDependency}[1]{
  \typeout{(#1)}
  \@addtofilelist{#1}
  \IfFileExists{#1}{}{\typeout{No file #1.}}
}
\makeatother

\newcommand*{\myexternaldocument}[1]{%
    \externaldocument{#1}%
    \addFileDependency{#1.tex}%
    \addFileDependency{#1.aux}%
}

\myexternaldocument{supplementary}

\usepackage{algorithm}
\usepackage{algpseudocode}

\usepackage{xcolor, comment}

\newcommand{\rev}[1]{{\color{black} {#1}}} 

\title{\rev{Political Context of the European Vaccine Debate on Twitter}}

\author[1,2]{Giordano Paoletti}
\author[1]{Lorenzo Dall'Amico}
\author[1]{Kyriaki Kalimeri}
\author[3,4]{Jacopo Lenti}
\author[1,*]{Yelena Mejova}
\author[1]{Daniela Paolotti}
\author[3,5]{Michele Starnini}
\author[1]{Michele Tizzani}
\affil[1]{ISI Foundation, Turin, Italy}
\affil[2]{Department of Control and Computer Engineering, Politecnico di Torino, 10129 Turin, Italy}
\affil[3]{CENTAI, Turin, Italy}
\affil[4]{Department of Computer, Control, and Management Engineering Antonio Ruberti, Sapienza University of Rome, Rome, Italy}
\affil[5]{Departament de F\'isica, Universitat Polit\`ecnica de Catalunya, Campus Nord, 08034 Barcelona, Spain}

\affil[*]{yelenamejova@acm.org}

\begin{abstract}
At the beginning of the COVID-19 pandemic, fears grew that making vaccination a political (instead of public health) issue may impact the efficacy of this life-saving intervention, spurring the spread of vaccine-hesitant content.
In this study, we examine whether there is a relationship between the political interest of social media users and their exposure to vaccine-hesitant content on Twitter.
We focus on 17 European countries using a multilingual, longitudinal dataset of tweets spanning the period before COVID, up to the vaccine roll-out.
We find that, in most countries, users' \rev{endorsement of} vaccine-hesitant content is the highest in the early months of the pandemic, around the time of greatest scientific uncertainty.
Further, users who follow politicians from right-wing parties, and those associated with authoritarian or anti-EU stances are more likely to \rev{endorse} vaccine-hesitant content, whereas those following left-wing politicians, more pro-EU or liberal parties, are less likely.
Somewhat surprisingly, politicians did not play an outsized role in the vaccine debates of their countries, receiving a similar number of retweets as other similarly popular users.
This systematic, multi-country, longitudinal investigation of the connection of politics with vaccine hesitancy has important implications for public health policy and communication. 
\end{abstract}

\begin{document}

\flushbottom
\maketitle
%
%
\thispagestyle{empty}

\section*{Introduction}
Despite the success of vaccination in reducing mortality and eradicating diseases like smallpox \cite{centers1999impact}, skepticism about vaccine safety and efficacy has persisted throughout history. 
The rapid development and global distribution of the COVID-19 vaccines spurred renewed apprehension.
In February 2022, a Eurobarometer survey found that while most EU citizens supported vaccination, concerns about unknown long-term side effects 
of COVID-19 vaccines ranged from 47\% to 70\% across different countries, with those who were altogether against vaccination reaching 29\% in Bulgaria, followed by 24\% in Slovakia and 21\% in Slovenia\cite{eurobarometer2022attitudes}.
Even before COVID-19, the World Health Organization has indicated vaccine hesitancy as one of the top 10 threats to global health in 2019 \cite{who2019ten}.


Vaccine hesitancy was found to correlate with a complex combination of psychological and sociological factors.
Recent literature has linked such attitudes with alternative health practices \cite{Kalimeri2019}, science denial \cite{browne2015going}, and conspiratorial thinking \cite{jolley2014effects}.
However, this must be contextualized in the personal experiences and beliefs, and in the broader societal, public health and communication environments \cite{dube2013vaccine}.
A concerning trend around web-based communication channels, including social networks and those integrating recommendation systems, is the possible formation of echo-chambers, both at national \cite{cossard2020falling,crupi2022echoes} and global scales \cite{lenti2022global}, as the tightly-knit, homogeneous communities in such echo-chambers provide a fertile ground for fringe narratives that oppose the mainstream \cite{monsted2022characterizing}.
These narratives often exclude traditional and authoritative sources \cite{murphy2021psychological} and instead are supported by low-quality information and misinformation~\cite{jennings2021lack}, which undermines the trust in the public health authorities. 




As governments around the world rushed to confront the COVID-19 pandemic, various political actors joined the discussion. 
At the same time, the increasing adoption of social media has coincided with its use by populist and anti-establishment politicians \cite{esser201728}, who took advantage of the context collapse around bite-size units of communication to promote skepticism of authority \cite{guerrero2020social}.
Worldwide, vaccine hesitancy has been linked to political beliefs, including in France and Italy \cite{peretti2020future, kreps2023resistance}, where those backing right-wing parties had a higher unvaccinated rate.
In Poland, around August 2021 vaccination rates were highest in the areas supporting the politician opposing the ruling conservative Law and Justice (PiS) party -- a party that has been accused of ``flirting'' with ``anti-vaxxers'' \cite{politico2021poland}. 
In the U.S., by May 2022 Republicans and Republican-leaning independents (60\%) were less likely than Democrats and Democratic leaners (85\%) to be fully vaccinated \cite{pew2022americans}. 
Overall, studies find vaccine hesitancy and political populism are positively associated across Europe \cite{recio2021vaccine,stoeckel2022politics}.


Although the connection between the use of social media and vaccine hesitancy has been documented \cite{jennings2021lack,clark2022role}, little attention has been paid to the connection between the politicized communication around vaccination on social media and vaccine hesitancy. 
\rev{For instance, some studies on Twitter have found politics to be a part of the conversation: COVID-19 vaccine-related discussions in Japan included criticism of the government's handling of the vaccine rollout and the holding of the 2020 Olympic Games in Tokyo \cite{kobayashi2022evolution}.}
As politicians take actions impacting \rev{public health and} the practice of medicine, including communicating to their constituents on the matters of personal health \cite{baron2022politicians}, it is urgent to understand the interplay between such political actors and their audience in the context of vaccination.


To address this research gap, we turn to one of the most popular social media platforms -- Twitter -- to gauge the relationship between political actors, political interest, and the consumption of vaccine-hesitant content in the period immediately before and during the onset of the COVID-19 epidemic. 
We propose a custom network analysis pipeline to identify users likely to \rev{retweet} to vaccine-hesitant content in 17 European countries using a multilingual, longitudinal dataset. 
Our approach extends previous research \cite{garimella2018quantifying} that assumes a two-sided controversy by also handling multiple ``pockets'' of opinion stances, making it generalizable to multi-sided discourse around the world.

Using these tools, we answer the following research questions:

\begin{description}[topsep=1pt, partopsep=0pt, itemsep=0pt, parsep=1pt]
    \item[RQ1] Are those more \rev{likely to endorse} vaccine-hesitant content interested in specific political parties?
    \item[RQ2] Is a user's politicization (\emph{i.e.} the extent of the interest in politics and the focus on few parties) related to their \rev{endorsement of} vaccine-hesitant content?
    \item[RQ3] Are political actors more influential in the vaccination debate compared to non-political users?
\end{description}

Crucially, our multilingual dataset allows us to examine the national discussions in the native languages, as out of the 17 countries considered, 16 do not have English as an official language (unlike in previous studies that use English-only queries to analyze ``worldwide'' vaccine sentiment \cite{ansari2021worldwide,reshi2022covid}).
Further, we perform an extensive mapping between the Twitter accounts in these countries and their respective political figures, which we in turn enrich using \textit{ParlGov}~\cite{DVN/UKILBE_2022}, a political science resource that provides detailed information about political parties, elections, cabinets, and governments in parliamentary democracies worldwide.
The combination of these resources has allowed us to complete the first quantitative investigation of the relationship between political interest on social media and vaccine hesitancy on a European scale, as detailed below.



\section*{Results}
\begin{figure*}[t]
\centering
\includegraphics[width=0.9\linewidth]{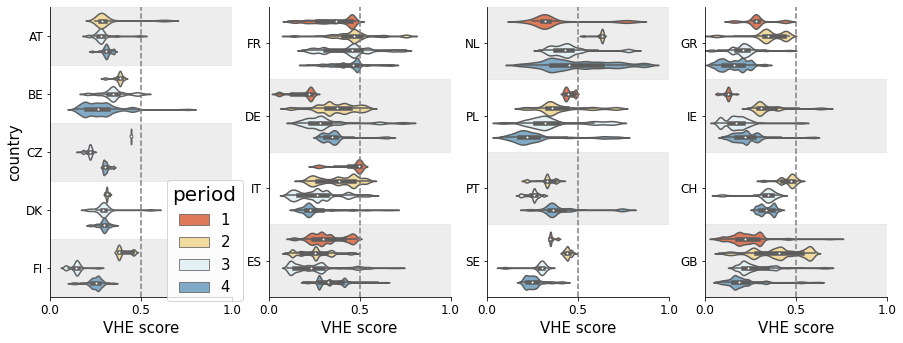}
\caption{\rev{Vaccine Hesitancy Endorsement (VHE)} score \rev{(see Algorithm~\ref{alg:VHE_score})} distribution across countries and periods. Dashed grey lines indicate scores where a user's exposure has equal shares of vaccine-hesitant and pro-vaccine content. Score near 1 indicates more hesitant, and 0 -- more pro-vaccine. \rev{Country abbreviations are ISO 3166 standard: AT \textit{Austria}, BE \textit{Belgium}, CZ \textit{Czech Republic}, DK \textit{Denmark}, FI \textit{Finland}, FR \textit{France}, DE \textit{Germany}, GR \textit{Greece}, IE \textit{Ireland}, IT \textit{Italy}, NL \textit{Netherlands}, PL \textit{Poland}, PT \textit{Portugal}, SE \textit{Sweden}, ES \textit{Spain}, CH \textit{Switzerland}, GB \textit{United Kingdom}.}}
\label{fig:VHE_score_violinplot}
\end{figure*}   

In this study, we focus on the Twitter debates around vaccination immediately before and during the COVID-19 pandemic in 17 European countries, in their official languages.
We begin by assigning each user in the vaccine debate a score that we call \rev{Vaccine Hesitancy Endorsement (VHE), spanning from $0$ to $1$ that gives an estimate of how likely the individual is to endorse a vaccine hesitant stance (VHE = $1$). 
The score is computed by combining label propagation and community detection on the country endorsement (retweet) network. 
These networks are obtained for $18$ European countries and are composed of posts published in the period 2019/10/01 - 2021/3/31. A number of tweets was manually annotated (see Methods for details) as ``vaccine-hesitant'', ``pro-vaccine'' or neutral.
We define vaccine-hesitant tweets as those stating directly the user will not vaccinate, questioning their efficacy or safety, or espousing conspiratorial views around their creation or distribution.
For each country and period, we then perform $100$ partial randomizations of the retweet network and run a community detection algorithm on each of the shuffled networks. Each community is assigned a score based on the proportion of the pro-, anti-vax and neutral labeled tweets. 
Finally the VSE score is assigned to each user by averaging the $100$ scores of the communities it belonged to in the different shufflings. 
Adding the shuffling step, we further quantify how tightly an individual is connected to a given community, allowing us to give a continuous score to each user.}

\rev{We begin by examining} the distributions of the VHE score for each country and one of four periods (\rev{the first before, while the remaining ones during COVID-19}), as shown in Figure \ref{fig:VHE_score_violinplot}. 
Note that some pairs country-period were excluded due to data sparsity.
One can see that in most countries and periods, VHE distributions are closer to zero (pro-vaccine) than 1 (vaccine-hesitant), indicating that the majority of users are more likely to \rev{endorse} pro-vaccine content.
Notable exceptions are France, with VHE distributions centered around 0.5 in all periods, the Netherlands, with broad VHE distributions, and Poland, where the majority of users have a VHE score below 0.5, but there exists a non-negligible minority of users \rev{tend to} anti-vaccine content.
Moreover, we can find interesting trends within each country, across time periods.
For instance, the distribution of VHE scores in Italy begins with most users \rev{likely to endorse} roughly the same amount of vaccine-hesitant content as the pro-vaccine one.
This trend then shifts towards 0, where more pro-vaccine content is easier to encounter for most users.
On the other hand, Germany starts out with most users having VHE scores close to zero, and over time develops a minority of users with much higher scores.
Yet in other cases, such as in France, the peak of the overall distribution remains stable over time.
In aggregate, however, the peak of the VHE score comes in period 2, during the early days of the pandemic, with a macro-average of 0.39; the VHE score goes down to 0.30 by period 4, during the vaccine rollout.


\medskip

\textbf{RQ1}. \emph{Are those more \rev{likely to endorse} hesitant content interested in specific political parties?}

To answer this question, recall that we model the vaccination debate in each country as a country-specific retweet network, one for each of the four time periods; for this analysis we choose only those that have at least 300 users (61 in total). 
For each country/period combination, we then perform an OLS regression to model a user's \rev{Vaccine Hesitancy Endorsement (VHE)} score using the user's followership of politicians in different parties as predictors (as well as some control variables, see Methods).
Out of these, 72\% (44) had an Adjusted R$^2$ score greater than or equal to 0.1, which we select for further analysis.
In these models, 266 coefficients of parties (51.6\%) were significant: 126 positive (having a positive relationship to the VHE score) and 140 negative,  where positive (negative) coefficients indicate that users who follow these parties are more (less) likely to \rev{endorse} vaccine-hesitant rather than pro content on Twitter.
For example, the most positive score was found by \textit{Alternative for Germany} in period 3, whereas the most negative by \textit{Civic Platform} in Poland (period 1) (for full listing of coefficients for each party across periods, see Data availability section).
Thus, we find mixed results -- the relationship between party interest and VHE score may vary between party, country, and period.

\begin{figure}[tbp]
\centering
   \centering  \includegraphics[width=0.5\linewidth]{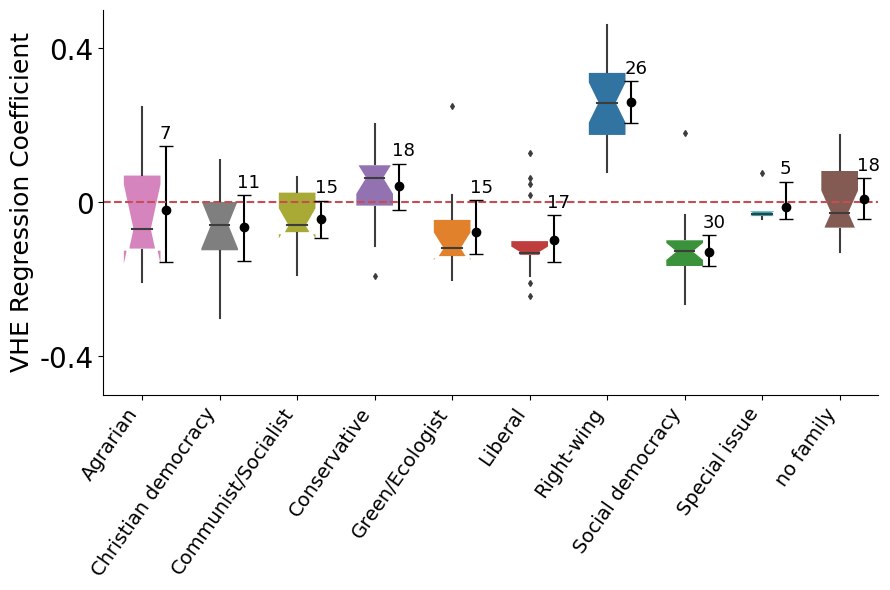}
    \caption{Distribution (boxplots) of OLS coefficients modeling users' \rev{Vaccine Hesitancy Endorsement (VSE)} score \rev{(see Algorithm~\ref{alg:VHE_score})} by their interest in parties, grouped by families using ParlGov\rev{, an extensive resource on political parties in parliamentary democracies}. 
    \rev{Positive (negative) coefficients indicate that users who follow these parties are more (less) likely to \rev{engage with} vaccine-hesitant rather than pro-vaccine content on Twitter}.
    Accompanying points and whiskers indicate a 99\% bootstrapped confidence interval. Numbers indicate how many parties are in each group.
    \rev{The parties identified as Right-wing have the strongest, and most positive, relationship with the VHE score.}}
\label{fig:party_family}
\end{figure}

\begin{figure*}[t]
\centering
  \includegraphics[width=0.90\linewidth]{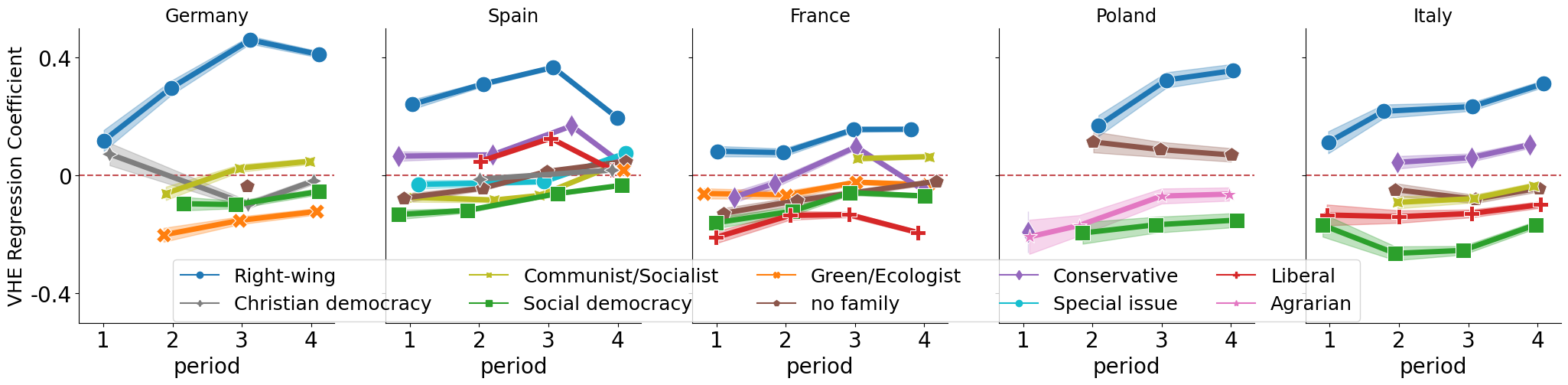}
\caption{Significant OLS coefficients (at $p<0.01$ with Bonferroni correction) \rev{modeling the users' VHE score using user interest in political parties}, grouped in families using ParlGoV. \rev{Shaded areas indicate the} 99\% confidence intervals. Showing countries having sufficient model fit over 4 time periods. \rev{For the most part, we find the effect sign of a political party family to remain consistent over time.}}
\label{fig:stability_trends}
\end{figure*}

\begin{figure*}[t]
  \centering
  \includegraphics[width=0.85\linewidth]{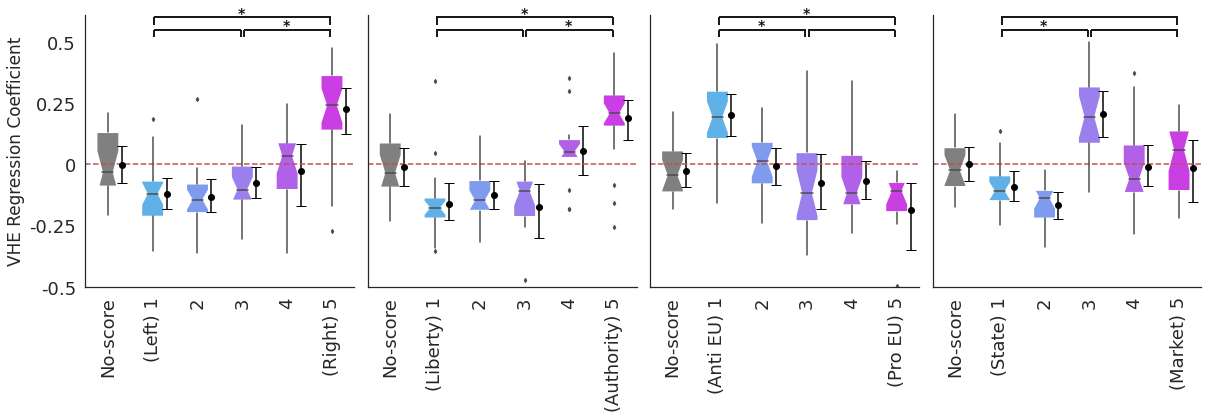}
\caption{Distribution (boxplots) of OLS coefficients modeling users' VHE score by their interest in parties having one of four dimensions defined by ParlGov, grouped in quintiles. Accompanying points and whiskers indicate a 99\% bootstrapped confidence interval. Horizontal brackets on top indicate the comparisons among quintiles $1$, $3$ and $5$, * signifies whether one distribution is statistically greater than the other (one-sided Mann-Whitney U test at $p<0.01$ with Bonferroni correction).}
\label{fig:fourdimensions}
\end{figure*}

To see if these findings generalize across countries, we group the parties according to the ParlGov classification (an extensive resource on political parties in parliamentary democracies\cite{DVN/UKILBE_2022}).
Figure \ref{fig:party_family} shows the distribution of coefficients for user interest in parties grouped in families using ParlGov with the accompanying 99\% bootstrapped confidence intervals.
We find that the parties identified as Right-wing have the strongest, and most positive, relationship with the VHE score. 
On the other hand, following parties in the Social democracy and Liberal families have a negative relationship with users \rev{endorsing} such vaccine-hesitant content.
We find no other statistically robust relationships for other party groups.

To check how consistent these results are over time, in Figure \ref{fig:stability_trends} we plot the significant coefficients and their confidence intervals for five countries that have models with Adjusted $R^2>0.1$ for all four periods.
We find that the sign of the coefficients rarely flips. 
The coefficients for the Right-wing family of parties remain positive, and that for Social Democracy remains negative. 
However, we find country-specific peculiarities: Liberal parties are associated positively with VHE score in Spain (and not in France or Italy), and the Green parties are more negatively in Germany.

Finally, we investigate the relationship between the VHE score and the four party characteristics as defined by ParlGov, namely Left vs.~Right, Liberty vs.~Authority, Anti vs.~Pro-EU, and State vs.~Market.
Figure \ref{fig:fourdimensions} shows the distribution of coefficients for parties in different quintiles of characteristics, accompanying bootstrapped confidence intervals, and horizontal brackets indicating statistical comparison using the one-sided Mann-Whitney U test.
For Left vs.~Right, Liberty vs.~Authority, and Anti vs.~Pro-EU dimensions, we find statistically robust differences between parties in the first and last quintiles, and sometimes with the middle quintile as well.
Users following Left-leaning politicians are less likely to \rev{endorse} vaccine-hesitant content, whereas those following Right-wing politicians are much more.
Similarly, those following parties closer to the Liberty characteristics -- those promoting expanded personal freedoms such as abortion, same-sex marriage, or greater democratic participation -- are less likely to \rev{endorse} vaccine-hesitant content, and the opposite for the parties closer to the Authority side -- those promoting authoritarian ideals of order, tradition, and stability.
Interestingly, even the party's stance on European Union correlates with the VHE score of users following them: those opposing European Union are more likely to have a higher VHE score.
On the other hand, the trend for the economic dimension of State (State-controlled economy) vs.~Market (free-market economy) displays a peak in association with the VHE score in the third quintile.

\medskip

\textbf{RQ2}. \emph{Is a user's politicization related to their \rev{endorsement of} hesitant content?}

\begin{figure} [t]
\centering
\includegraphics[width=0.98\linewidth]{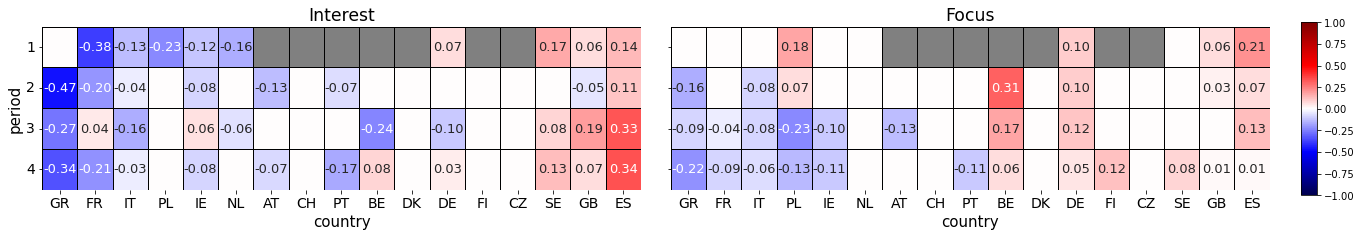}
\caption{\rev{Spearman correlation between political interest and VHE score (left) and political focus and VHE score (right) (see RQ2 Methods). Grey cells indicate country/periods having fewer than 300 users. White cells indicate a non-significant correlation.}
}
\label{fig:Interest_Focus_Heatmap}
\end{figure}

We address this question by defining two measures of politicization for each user: \emph{political interest} (the proportion of all accounts a user follows that are politicians) and \emph{political focus} (the share of politicians in the user's most followed party). 
Figure \ref{fig:Interest_Focus_Heatmap} shows the Spearman rank correlation coefficient between these two measures and the VHE score, for each country/period (those that have fewer than 300 users are greyed out).
First, note that the two measures tend to produce similar results for the same countries.
Second, the relationship may be different for different countries.
For some, there is a positive relationship between both measures and VHE, especially Spain, indicating that (in the case of interest) the greater the share of politicians that a user follows, the more likely they are to \rev{endorse} vaccine-hesitant content in their timeline.
On the other hand, other countries display a negative correlation, especially Greece, suggesting that a greater political interest corresponds to less vaccine-hesitant content.
However, this relationship tends to remain constant over time, though can sometimes change, as in the case of Poland when considering political focus. 
Thus we conclude that the relationship between politicization and VHE is not straightforward, but specific to a particular country and its political situation (echoing our findings in RQ1).

\medskip

\textbf{RQ3}. \emph{Are political actors more influential in the vaccination debate compared to non-political users?}

Finally, we compare the engagement with content posted by political actors ---the Twitter accounts of prominent politicians--- to other users in that country.
We match politicians with other users on the number of followers, followees, and daily posting rate in order to achieve a fair comparison.
We then consider country/period networks in which there are at least 10 politicians involved in the retweet network, resulting in 44 experiments.
First, when comparing the number of retweets these users have received, we find that 40 out of 44 comparisons (91\%) do not show a significant difference (using one-sided Wilcoxon test, with Bonferroni correction); only in 4 cases political accounts had more retweets.
The result is the same if we consider the number of unique retweeters. 
Second, measuring the PageRank centrality of both groups, we find only 2 cases when they are significantly different (politicians having higher centrality).
Third, considering mentions, we find that in 10 cases (less than a quarter of all cases) politicians are mentioned significantly more than others, but there are no cases of the reverse.
In summary, we do not find evidence of political Twitter users being much more influential than other users who have a similar social profile, using the metrics above.

\section*{Discussion}


A connection between far-right and anti-vaccination movements has been seen on the streets of many European countries, with documented ``anti-vax'' protests of white supremacists in Britain \cite{saphore2021white} and arrests of Italy’s far-right New Force during COVID-19 riots \cite{broderick2021italy}.
However, our study shows that the connection between interest in right-wing political actors and vaccination-hesitant content is not limited to the extreme cases on the streets, but can be found on one of the largest social media websites. 
We notice that, according to ParlGov, during the study period no right-wing government was in power in the considered countries (with the exception of North League from February 2021 in Italy), thus they were in an opposition role, which may have shaped their audience.

Our findings are further supported by survey evidence.
A large survey conducted from December 2020 to January 2022 in Spain has shown that far-right supporters were almost twice as likely to be vaccine-hesitant than the overall population -- a consistent trend with a brief lessening around October 2021 \cite{serrano2023far}.
\rev{A survey of UK residents in April-May, 2020 found that those believing in conspiracy theories around the ``authorities'' exaggerating COVID-19 deaths and pushing vaccination, among other conspiratorial beliefs, corresponded with a lack of engagement in health-protective behaviours~\cite{allington2021health}.}
Elsewhere, a survey of the residents of Norway showed that the refusal to vaccinate is associated with right-wing ideological constraint, with the authors concluding that "vaccine refusal is partly an act of political protest and defiance which attaches itself to consistent right-wing attitudes'' \cite{wollebaek2022right} \rev{(a similar association between right-wing political attitude and refusal to vaccinate was found in a German survey~\cite{fischer2023metacognition})}.
However, a survey of French respondents found that those both on the right and left extremes were more likely to refuse vaccination \cite{peretti2020future}.
A more refined look at the personal right-wing ideology was used in a survey of respondents in Germany, Poland, and the United Kingdom.
When separated into two distinct right-wing dimensions, high ``social dominance orientation'' (SDO) is associated with higher vaccine hesitancy, while high ``right-wing authoritarianism'' (RWA) is associated with lower hesitancy, pointing to additional nuance within political ideologies~\cite{bilewicz2022politics}. 
As we show, political situations and their association with vaccination hesitancy vary greatly amongst countries, more work is necessary to understand the local political incentives around public health communication. 

These findings, however, also present an opportunity.
The fact that those who are likely to find \rev{and endorse} vaccine-hesitant information on social media are likely to follow particular political actors points to a possible direct way to communicate with them.
If such actors can be persuaded to team up with the public health authorities to promote scientifically-grounded information, their audience would receive such messages from sources they already trust.
For instance, the partisan divide in terms of COVID vaccination in the UK has been shown to be less drastic than in the US, likely to the greater pro-vaccination position of the Conservative UK government \cite{klymak2022partisanship}.
Additionally, public health messaging and interventions could be tailored to specific political affiliations or beliefs to better address vaccine hesitancy.

Note that, even though we did not explicitly label content for misinformation, labelers have encountered many instances of possible misleading or erroneous information \rev{(which has been found to be posted even by accounts identified as ``Business/NGO/government''~\cite{kouzy2020coronavirus})}. 
Further, we acknowledge that our study does not include bot detection, as bots may play an important role in information propagation~\cite{gallotti2020assessing}.
A recent study has shown there is a global proliferation of low-quality information within and across national borders \cite{lenti2022global}. 
Whether this information is propagated by political actors is an important question in terms of accountability and responsibility of those representing a public office.
For instance, in early 2023, the UK Conservatives suspend a lawmaker for posting vaccine misinformation on Twitter \cite{ap2023uk}.







This study has certain notable limitations.
Although the focus of this paper is Europe, by far not all countries were included, either due to data sparsity (despite the multi-lingual queries) or the lack of external resources (such as ParlGov).
Twitter users are not representative of the larger populations and tend to be more political \cite{bestvater2022politics}, potentially overestimating political engagement in the whole country. 
Although we find some relationships which are stable over time, both public health and political spheres are highly dynamic, limiting our findings to the unique (and unprecedented) time of the rollout of COVID-19 vaccines.
Further, keyword-based data collection is always bound to miss content that is phrased differently from the query, with additional challenges in a multilingual setting. 
However, our decision to keep the keyword set consistent across countries and time was to conserve comparability between countries and to avoid topic drift (for instance, discussion around vaccine-related ``green passes'' during the vaccination period).
The methodology used in this study also favors users who are active and engage in retweeting, excluding those who engage with the topic rarely (previous work shows it is difficult to ascertain the opinions of such users even for human annotators \cite{cossard2020falling}).
Observational studies also fail to capture the impact of merely seeing information, that is, whether encountering vaccine-hesitant content actually changes people's minds, although previous studies have found a link between social media use and COVID-19 vaccine hesitancy \cite{vaccines9060593,sallam2021high}.  
Complementary methodologies including surveys may be needed to further understand cognitive and psychological nuances of the interaction between social media use, political interest, and vaccine stance deliberation.




Finally, as this study deals with healthcare decisions that may be highly personal in nature, privacy is an important consideration of this work.
Because the Streaming API was used to collect this data, it is possible that some of the content was removed either by the user or by Twitter by the time of the analysis.
However, besides annotating select tweets, throughout the paper we deal with the structural properties of the RT network and followership of popular political accounts, and present our findings in an aggregated fashion.
Further, our methodology identifies pockets of users agreeing on a topic, which could be misused to target or harass people because of their opinions. 
Any interventions, therefore, possibly based on this analysis must comply with the ethical standards of public communication, and further research possibly dealing directly with individuals must follow the human research guidelines of their institution \cite{humansubjectsresearch}.

\section*{Methods}

Below, we describe the initial data collection using the Twitter Streaming API, the selection of country-specific content using user geo-location, and the construction of retweet networks that represent one user's endorsement of another's content.
We then enrich this data from two sides: by estimating the vaccination stance of the tweets users are likely to encounter in these networks, and by estimating the political interest of these users.
Finally, we compare these two sides to answer our research questions.

\subsection*{Data Collection} 

First, in 2019 we compiled a catalog of terms related to vaccines that are translated into 18 different languages (using 2-letter ISO: \emph{bg, cz, fi, de, el, en, es, fi, fr, hu, it, nl, pl, pt, ro, ru, sv, tr}) starting from the list created for the research of tracking vaccination discussion on Twitter~\cite{cossard2020falling}.
Using Twitter search, we iteratively added keywords to the list and searched again until no new keywords could be found. 
Then, these keywords were translated by native speakers with the task of including several common grammatical variations or relevant local keywords. 
This resulted in a collection of 459 keywords, including \emph{vaccine}, \emph{novax}, \emph{measles}, \emph{MMR}, \emph{vaccinated}, and others (for full listing, see Data availability). 
Once the collection began, we did not modify the list to incorporate new vaccine-related words to keep the selection process consistent and the data comparable across time. 
This methodological choice likely led to some exclusion of the tweets mentioning new vaccines that did not also use the above keywords (see sec. Limitations). 


The final collection, spanning 2019/10/01 - 2021/3/31, is divided into four three-month periods: (1) {pre-COVID} (October 2019 - December 2019), (2) {pre-vaccine} (July 2020 - September 2020), (3) {vaccine development} (October 2020 - December 2020), and (4) {vaccine rollout} (January 2021 - March 2021) periods (see Supplementary Figure S\ref{fig:tweet_per_day} for a volume graph).
The time periods were chosen based on the international news about the vaccine development, though we acknowledge that they fit loosely the vaccine rollout schedule in each individual country.
For instance, in period 2 the Sputnik V vaccine was announced (2020/08/11), in period 3 Pfizer-BioNTech (2020/11/09) and Moderna (2020/12/18) vaccines were announced, and at the beginning of period 4, the first AstraZeneca vaccine was administered (2021/01/04).
Note that we exclude the period from January to June 2020, when the COVID pandemic first begins.
This results in a dataset of $319M$ of tweets.
The distribution across various languages exhibits a marked heterogeneity, with English accounting for 49.0\% of the tweets, followed by Spanish at 24.8\%, Portuguese at 13.7\%, and French at 4.7\%. 
Other languages, including Czech, Finnish, Danish, Romanian, Bulgarian, and Hungarian, each represent a negligible portion of the traffic, constituting less than 0.1\% of the total (note that they were queried separately, so the large volume of popular languages should not have affected their collection).

\paragraph{Geo-localization}

Next, we assigned to each tweet a location using the self-reported location field provided by users in their description, by matching it to GeoNames\cite{geonames}, a large geographical database of locations. 
We manually eliminated over 500 words most commonly associated with non-locations for false matches. 
Using this approach, we were able to geolocate more than 49\% of the users. 
Since some users may provide a fake position or matching may fail (e.g. in the case of homonym places), to reinforce proper geo-localisation, for each country we filtered out tweets written in a language other than the official ones of their country. 
We tested this approach using tweets that come with geographic coordinates in their metadata as the gold standard, resulting in 93.7\% (95\% CI [83.2, 100.]) accuracy (which is on average more than twice as accurate as localizing users writing in other languages).  See Supplementary Table S\ref{tab:accuracy_of_geolocalization} for accuracy estimates by country by language.
Finally, we constrained our analysis to the European countries appearing in the ParlGov dataset (more on this below in the Political Analysis section).

\paragraph{Debate as networks}

Following previous literature \cite{cossard2020falling,crupi2022echoes,lenti2022global}, we represent the vaccine debate in each country using directed, weighted graphs that capture retweet interactions among users. 
It has been widely shown (e.g. \cite{conover2011political}) that retweet networks have high homophily concerning intrinsic characteristics of the nodes, such that users who retweet each other are likely to share the same opinions and are exposed to the same type of content\rev{, thus such networks are often called \emph{endorsement networks}\cite{garimella2017effect}}. 
For each country and period, we built a retweet (RT) network where the nodes are the users in the dataset and the weight on edges from node \emph{i} to \emph{j} is the number of times user \emph{i} retweeted user \emph{j}'s tweets. 
For each network, we kept the nodes in the biggest \emph{weakly connected component} (WCC), which on average corresponds to 93\% of the total (while the size of the second largest WCC is usually less than 1\%.). 
In the following analysis, we considered RT networks with at least 300 nodes from the following 17 countries: Austria, Belgium, Czech Republic, Denmark, Finland, France, Germany, Greece, Ireland, Italy, Netherlands, Poland, Portugal, Spain, Sweden, Switzerland, and United Kingdom.
This resulted in 61 networks (17 countries $\times$ 4 periods), as 7 networks were removed from period 1 due to the size threshold. In the Supplementary Table S\ref{tab:officiallanguages} the sizes of the WCCs in each country by period are listed.

\subsection*{Vaccination Stance Labeling}

\begin{algorithm}[!t]
\caption{Assigning users' VHE score}\label{alg:VHE_score}
\begin{algorithmic}[1]
    \State $G \gets G(V,E)$ \Comment{RT-Network G}
    \For{$i \gets 1$ to $100$}
        \State $G' \gets$ Perturb the network $G$
        \State $\{c\} \gets$ Community detection on $G'$
        \State $\gamma^{(c)}_{VH} \gets$ VH Exposure in each community $c$  
    \EndFor
    \For{$u \in V$} \Comment{V nodes of G}
        \State \emph{VHE}$_u \gets$ Average of $\gamma^{(c)}_{VH}$ where $c$'s are the communities to which $u$ was assigned. 
    \EndFor
  \end{algorithmic}
\end{algorithm}

Next, we turn to the labeling of the tweets in terms of vaccination stance (and measuring users' exposure to these stances).
To this end, we create a dataset of manually annotated samples of the content shared on the RT networks. 
To avoid the over-representation of some portions of the social network, we performed a stratified sampling of the tweets first dividing the networks into $15$ communities and then selecting $6$ tweets per community with the largest difference between the fraction of internal and external retweeters (i.e. the fraction of users from inside a community retweeting the post, minus a fraction of users retweeting it outside). 
The chosen number of communities and tweets per community are a trade-off between an accurate representation and our capability of manual annotation. 
Notably, in this step, community detection serves to perform a stratification, this is why the number of communities is the same for all networks.
This strategy allowed us to identify tweets that are at the same time popular and representative of a network portion.
In this way, we were able to achieve a high coverage ranging between 37.5\% (1st period) and 4.5\% (4th period) of the total number of tweets produced by users in the RT network and an average fraction of users covered (i.e. for whom we noted at least one of their posts) between 61.8\% to 38.5\%.

The selected tweets were then labeled by 13 expert annotators with a background in the vaccine debate and as far as possible proficient in the original language of the tweets. 
Nevertheless, we translated all tweets into English using Google Translate for cross-checking. 
The classification task consisted in deciding whether a tweet was \emph{pro-vaccine}, \emph{vaccine-hesitant}, or \emph{other}. 
In particular, vaccine-hesitant tweets included those stating directly the user will not vaccinate, questioning their efficacy or safety, or espousing conspiratorial views around their creation or distribution.
The annotators were encouraged to take into account all information available (images, videos, URLs, etc.) if the tweet was still available online. 
Annotator agreement, measured using \emph{Cohen's kappa} of $k=0.47$ showed the task to be moderately complex.
However, when the label \emph{other} is excluded, the agreement is $k=0.78$, with an overlap in labels of 92.3\%.
We make the labeled dataset of 5667 tweets in 13 languages available to the research community (see Data availability).




With a partial annotation of the tweets at hand, we aim at assigning to each user a score, which we dub \emph{Vaccine Hesitancy Exposure} (VHE) score, capturing the stance of the content they might be exposed to
but not necessarily their own stance on vaccination. 
To accomplish this task, we begin by propagating the labels of the manually annotated data to other users who have retweeted (but not ``quoted'') them. 
Next, we apply a procedure summarized in Algorithm \ref{alg:VHE_score}. 
In simple words, we randomize the network, perform community detection and assign a score in [0,1] to each individual to be exposed to vaccine-hesitant content, given its class affiliation. The process is repeated $100$ times, and the VHE score is obtained by averaging the result over all trials.
We now describe in detail the steps in the scheme of the algorithm.

\paragraph{Perturb the network:} We generate 100 versions of the network using perturbation methods described in \cite{casas2017survey}. In this way, we attempt to mitigate weaker clustering signals that may be an artifact of the particular sample of the RT network.
We sample randomly 15\% of the retweets and change their target, selecting the new one according to the \textit{weighted-in-degree} distribution of the nodes, such that the account popularity information is preserved.
We empirically choose a fraction of 15\% as the maximum amount of noise we can introduce before the smallest networks' community structure is unrecoverable, while still allowing us to have significant randomization effects on the denser networks of the largest countries.

\paragraph{Community detection (CD):} Both for the stratified sampling and for the VHE definition, we perform community detection with the spectral clustering algorithm for weighted networks introduced in \cite{dall2021nishimori}, applied to a symmetrized version of our graph. We adopt the ``spin-glass'' version of the regularized Laplacian matrix and extend its use to more than two communities as per \cite{dall2021unified}. 
We choose this algorithm for its speed, its efficiency on weighted, and sparse graphs, but also because it is one of the few known methods to estimate the number of communities in graphs. 
For interpretation purposes, we split the graph into $k=15$ communities, whenever the estimated number exceeded this threshold. 
As a robustness check, we compared the VHE scores created using the Louvain algorithm \cite{blondel2008fast}, as well as the results for RQs 1 \& 2, and obtained generally consistent results to those using spectral clustering (for more on VHE score robustness checks, see Supplementary Section S\ref{SM:VHErobustness} and additional results in Supplementary Figures S\ref{fig:party_family_Louvain}-S\ref{fig:Interest_Focus_Heatmap_Louvain}). 


\paragraph{\rev{Stance Endorsement} within communities:} 
In each community $c$ detected above, we define a measure $\gamma^{(c)}_{VH}$ that captures the \rev{extent to which its members are likely to endorse vaccine-hesitant content, }
defined as the difference between the fraction of hesitant and pro tweets: 
    $$
    \gamma^{(c)}_{VH} = \frac{1}{2}\left[\frac{N_{VH}-N_{Pro}}{N_{VH}+N_{Pro}+N_{other}} + 1\right]
     $$
where $N_{VH}$, $N_{Pro}$, $N_{other}$ is the number of vaccine-hesitant, pro-vaccine, and other tweets in the community. The value $\gamma^{(c)}_{VH}$ spans between $0$ and $1$ and captures how much likely it is that a user in the community $c$ \rev{retweets} 
a hesitant tweet instead of a pro-vaccine one.

\paragraph{From community-level to user-level scores:}
Finally, for each user $u$ we compute their \rev{Vaccine Hesitance Endorsement} \emph{VHE}$_u$ score as the mean on the set of 100 values of $\gamma^{(c)}_{VH}$ where the $c$'s are the communities to which $u$ was assigned.
The process results in a score for each user from 0 to 1, where 0 means that \rev{the user is more likely to retweet a pro-vax tweet, and 1 more likely to retweet a vaccine-hesitant one.}
Intuitively, having a value greater than 0.5 means that on average the user was sorted into communities where more hesitant tweets were shared than pro-vaccine. 
\rev{As a validation of this approach, we consider the 4th time period (the one having most activity) and annotate a sample of 5 tweets for a sample of users stratified by VHE score in Italy, France, Spain, and United Kingdom, each country having between 225-275 tweets. 
The Spearman correlation between the the difference between anti- and pro-vax tweets for a user and their VHE score ranged from 0.38 to 0.57, and out of all anti-vax tweets, 50-83\% were posted by the users assigned VHE score in the highest tercile (for details, see Supplementary Section S\ref{SM:VHEvalidation}).}

\subsection*{Political Analysis}
\label{sec:politicalanalysis}

In order to study the connections between VHE score and politics, we identified and characterized the Twitter accounts in our dataset linked to politicians from each respective country.

\subsubsection*{Identifying Politicians \& Parties.} 
We began by building an extensive list of politicians (parliamentarians and parties' leaders) who are associated with a personal Twitter account using \textit{Politicians on Social Media} \cite{haman2021politicians} and \textit{Twitter Parliament} \cite{van2020twitter} datasets. 
Then, we annotated them with the party affiliations in the target period (October 2019 -- March 2021) by first matching them to \textit{WikiData} \cite{vrandevcic2014wikidata} and then manually annotating those not found in WikiData using all available resources.
Since parties may be dynamic (dissolution, merging, renaming, etc.) we took as gold standard those parties which are present in the national Parliament during the last government session up to December 2021 provided by the \textit{ParlGov} dataset (the \textit{Parliamentary Governments Database} \cite{DVN/UKILBE_2022} is an extensive and widely-utilized resource providing detailed information about political parties, elections, cabinets, and governments in parliamentary democracies worldwide). 
For those parties not in \textit{ParlGov} but which were active in the target period, we either labeled them with a larger party in \textit{ParlGov} that they have joined, or as \textit{Other} if a larger party could not be found.
Indeed, it is important to take smaller parties into account when analyzing social media data, since their supporters can be vocal on Twitter \cite{jungherr2012pirate}. A link to the list of all political users annotated is provided in the Data availability section.

\subsubsection*{Comparison of User Interest in Political Parties to the VHE Score (RQ1).}

To compare the user's political interests to their VHE score, we employ an OLS regression model that predicts a user's VHE score using the fraction of politicians followed by them in each party, along with a set of confounding variables.
These confounding variables include the number of followers and followees, daily posting rate, weighted in-degree and weighted out-degree (number of retweets they had, and number of retweets they made in the vaccine debate, respectively), and the proportion of followed users who are politicians (political interest, defined above).
These variables were selected using the Variance Inflation Factor (VIF), to make sure they do not introduce multicollinearity. 
We then standardized all features within each country, including the target variable.
For each country/period network, we run OLS and note the model fit (Adjusted $R^2$) and the coefficients of the politically-related variables.
We compute 61 such models, one for each network except for the 7 networks that have fewer than 300 users in the WCC. 
Further, we only consider the models which have a fit of Adjusted $R^2 > 0.1$ \cite{ozili2023acceptable}. 
To alleviate the multiple hypothesis testing problems, we apply the Bonferroni correction to the $p$-values of the variable coefficients, selecting those significant at $p < 0.01$, with the correction.
Finally, we report aggregated results in terms of the proportion of significant coefficients, their direction (positive vs.~negative), and the magnitude and stability of coefficients for select parties.

We perform a similar analysis by aggregating the parties by classification assigned by ParlGov in terms of the political family (conservative, social democrats, liberal, green, etc.), and four dimensions: left/right, State/market (economic policy), liberty/authority (personal freedom), pro/anti-EU. 
For instance, when aggregating per political family, we run a model for each network that models the VHE score by considering the proportion of politicians followed by the user in a particular political family, along with a set of confounding variables.
In the case of the four dimensions, each dimension is run as a separate model, with the numerical score of the dimension binned into quintiles (as well as a ``none" score when a party does not have a score in that dimension). 
For each dimension, the quintile bins were computed on the whole dataset, before being applied to the parties in each country.
Similar Adjusted $R^2$ and Bonferroni-corrected $p-$value filters were applied to these models.
When comparing the average coefficients of the variables in different groups, we compute a confidence interval (CI) using bootstrapping ($n=1000$).
\rev{See Supplementary Material Section S\ref{SM:RegressionDetails} for details on the features used in and performance of the above regressions.}

\subsubsection*{Comparison of User Politicization to the VHE Score (RQ2).} 

To assess the politicization of the users in our data, we collected the followers of the politician identified above 
using the Twitter Followers API.
The share of users who follow at least one politician varies on average between 54\% and 88\% except for Portugal at only 22\%. 
Note that, as the API does not provide historical knowledge of the follower relationship, here we assume that the followership does not drastically change over time (between the start of our data on October 2019 and the followership collection in December 2022).
We then define two measures of user politicization by each user: \emph{political interest} is the proportion of all accounts a user follows that are politicians, and \emph{political focus} is the share of politicians in the user's most followed party (for example, if a user follows 5 politicians in party A, 3 in party B, and 2 in party C, the political focus is 5/10 = 0.5).
\rev{To make sure the metrics are computed on enough information, we constrain our consideration to users who follow at least 5 politicians when computing political focus (excluding on average 44\% (99\% CI [20, 93]) users), and to users who follow at least 100 accounts (not necessarily politicians) when computing political interest (excluding on average 9\% (99\% CI [4, 18]) users).}
Finally, we compute the Spearman correlation between each of these two measures and the VHE score.



\subsubsection*{Comparison of Politicians to Others (RQ3).}

Finally, to ascertain the importance of politicians in the vaccination debates in each country, we compare the politicians and their posts to a matched set of other users in terms of retweets and mentions, and their PageRank in the RT network (using a one-sided Wilcoxon signed-rank test).
We also considered the unique sets of retweeters and those mentioning the politicians, getting the same results.
To perform a fair comparison, we match each political account with another on the number of followers, followees, and the daily posting rate using the Euclidean distance with the standardization of all variables.
The distributions of these variables with matches were checked with paired t-tests to make sure the political accounts were indeed close to the matched baseline.

\bibliography{polivax}

\begin{thebibliography}{10}
\urlstyle{rm}
\expandafter\ifx\csname url\endcsname\relax
  \def\url#1{\texttt{#1}}\fi
\expandafter\ifx\csname urlprefix\endcsname\relax\def\urlprefix{URL }\fi
\expandafter\ifx\csname doiprefix\endcsname\relax\def\doiprefix{DOI: }\fi
\providecommand{\bibinfo}[2]{#2}
\providecommand{\eprint}[2][]{\url{#2}}

\bibitem{centers1999impact}
\bibinfo{author}{{Centers for Disease Control and Prevention (CDC)}}.
\newblock \bibinfo{journal}{\bibinfo{title}{Impact of vaccines universally
  recommended for children--united states, 1990-1998}}.
\newblock {\emph{\JournalTitle{MMWR. Morbidity and mortality weekly report}}}
  \textbf{\bibinfo{volume}{48}}, \bibinfo{pages}{243--248}
  (\bibinfo{year}{1999}).

\bibitem{eurobarometer2022attitudes}
\bibinfo{author}{{Flash Eurobarometer 505}}.
\newblock \bibinfo{title}{Attitudes on vaccination against covid-19 - february
  2022}.
\newblock
  \bibinfo{howpublished}{https://www.quotidianosanita.it/allegati/allegato1650373320.pdf}
  (\bibinfo{year}{2022}).

\bibitem{who2019ten}
\bibinfo{author}{{World Health Organization}}.
\newblock \bibinfo{title}{Ten threats to global health 2019}.
\newblock
  \bibinfo{howpublished}{https://www.who.int/news-room/spotlight/ten-threats-to-global-health-in-2019}
  (\bibinfo{year}{2019}).

\bibitem{Kalimeri2019}
\bibinfo{author}{Kalimeri, K.} \emph{et~al.}
\newblock \bibinfo{title}{Human values and attitudes towards vaccination in
  social media}.
\newblock In \emph{\bibinfo{booktitle}{Companion Proceedings of The 2019 World
  Wide Web Conference}}, \bibinfo{pages}{248--254} (\bibinfo{year}{2019}).

\bibitem{browne2015going}
\bibinfo{author}{Browne, M.}, \bibinfo{author}{Thomson, P.},
  \bibinfo{author}{Rockloff, M.~J.} \& \bibinfo{author}{Pennycook, G.}
\newblock \bibinfo{journal}{\bibinfo{title}{Going against the herd:
  psychological and cultural factors underlying the ``vaccination confidence
  gap''}}.
\newblock {\emph{\JournalTitle{PLoS One}}} \textbf{\bibinfo{volume}{10}},
  \bibinfo{pages}{e0132562} (\bibinfo{year}{2015}).

\bibitem{jolley2014effects}
\bibinfo{author}{Jolley, D.} \& \bibinfo{author}{Douglas, K.~M.}
\newblock \bibinfo{journal}{\bibinfo{title}{The effects of anti-vaccine
  conspiracy theories on vaccination intentions}}.
\newblock {\emph{\JournalTitle{PloS one}}} \textbf{\bibinfo{volume}{9}},
  \bibinfo{pages}{e89177} (\bibinfo{year}{2014}).

\bibitem{dube2013vaccine}
\bibinfo{author}{Dub{\'e}, E.} \emph{et~al.}
\newblock \bibinfo{journal}{\bibinfo{title}{Vaccine hesitancy: an overview}}.
\newblock {\emph{\JournalTitle{Human vaccines \& immunotherapeutics}}}
  \textbf{\bibinfo{volume}{9}}, \bibinfo{pages}{1763--1773}
  (\bibinfo{year}{2013}).

\bibitem{cossard2020falling}
\bibinfo{author}{Cossard, A.} \emph{et~al.}
\newblock \bibinfo{title}{Falling into the echo chamber: the italian
  vaccination debate on twitter}.
\newblock In \emph{\bibinfo{booktitle}{Proceedings of the International AAAI
  conference on web and social media}}, vol.~\bibinfo{volume}{14},
  \bibinfo{pages}{130--140} (\bibinfo{year}{2020}).

\bibitem{crupi2022echoes}
\bibinfo{author}{Crupi, G.}, \bibinfo{author}{Mejova, Y.},
  \bibinfo{author}{Tizzani, M.}, \bibinfo{author}{Paolotti, D.} \&
  \bibinfo{author}{Panisson, A.}
\newblock \bibinfo{title}{Echoes through time: Evolution of the italian
  covid-19 vaccination debate}.
\newblock In \emph{\bibinfo{booktitle}{Proceedings of the International AAAI
  Conference on Web and Social Media}}, vol.~\bibinfo{volume}{16},
  \bibinfo{pages}{102--113} (\bibinfo{year}{2022}).

\bibitem{lenti2022global}
\bibinfo{author}{Lenti, J.} \emph{et~al.}
\newblock \bibinfo{journal}{\bibinfo{title}{Global misinformation spillovers in
  the online vaccination debate before and during covid-19}}.
\newblock {\emph{\JournalTitle{JMIR Infodemiology}}}  (\bibinfo{year}{2023}).

\bibitem{monsted2022characterizing}
\bibinfo{author}{M{\o}nsted, B.} \& \bibinfo{author}{Lehmann, S.}
\newblock \bibinfo{journal}{\bibinfo{title}{Characterizing polarization in
  online vaccine discourse—a large-scale study}}.
\newblock {\emph{\JournalTitle{PloS one}}} \textbf{\bibinfo{volume}{17}},
  \bibinfo{pages}{e0263746} (\bibinfo{year}{2022}).

\bibitem{murphy2021psychological}
\bibinfo{author}{Murphy, J.} \emph{et~al.}
\newblock \bibinfo{journal}{\bibinfo{title}{Psychological characteristics
  associated with covid-19 vaccine hesitancy and resistance in ireland and the
  united kingdom}}.
\newblock {\emph{\JournalTitle{Nature communications}}}
  \textbf{\bibinfo{volume}{12}}, \bibinfo{pages}{29} (\bibinfo{year}{2021}).

\bibitem{jennings2021lack}
\bibinfo{author}{Jennings, W.} \emph{et~al.}
\newblock \bibinfo{journal}{\bibinfo{title}{Lack of trust, conspiracy beliefs,
  and social media use predict covid-19 vaccine hesitancy}}.
\newblock {\emph{\JournalTitle{Vaccines}}} \textbf{\bibinfo{volume}{9}},
  \bibinfo{pages}{593} (\bibinfo{year}{2021}).

\bibitem{esser201728}
\bibinfo{author}{Esser, F.}, \bibinfo{author}{Stepi{\'n}ska, A.} \&
  \bibinfo{author}{Hopmann, D.~N.}
\newblock \bibinfo{journal}{\bibinfo{title}{28. populism and the media.
  cross-national findings and perspectives}}.
\newblock {\emph{\JournalTitle{T. Aalberg, F. Esser, C. Reinemann, J.
  Str{\"o}mb{\"a}ck \& C. d. Vreese (Eds.), Populist political communication in
  Europe}}} \bibinfo{pages}{365--380} (\bibinfo{year}{2017}).

\bibitem{guerrero2020social}
\bibinfo{author}{Guerrero-Sol{\'e}, F.}, \bibinfo{author}{Su{\'a}rez-Gonzalo,
  S.}, \bibinfo{author}{Rovira, C.} \& \bibinfo{author}{Codina, L.}
\newblock \bibinfo{journal}{\bibinfo{title}{Social media, context collapse and
  the future of data-driven populism}}.
\newblock {\emph{\JournalTitle{Profesional de la informaci{\'o}n}}}
  \textbf{\bibinfo{volume}{29}} (\bibinfo{year}{2020}).

\bibitem{peretti2020future}
\bibinfo{author}{Peretti-Watel, P.} \emph{et~al.}
\newblock \bibinfo{journal}{\bibinfo{title}{A future vaccination campaign
  against covid-19 at risk of vaccine hesitancy and politicisation}}.
\newblock {\emph{\JournalTitle{The Lancet infectious diseases}}}
  \textbf{\bibinfo{volume}{20}}, \bibinfo{pages}{769--770}
  (\bibinfo{year}{2020}).

\bibitem{kreps2023resistance}
\bibinfo{author}{Kreps, S.~E.} \& \bibinfo{author}{Kriner, D.~L.}
\newblock \bibinfo{journal}{\bibinfo{title}{Resistance to covid-19 vaccination
  and the social contract: evidence from italy}}.
\newblock {\emph{\JournalTitle{npj Vaccines}}} \textbf{\bibinfo{volume}{8}},
  \bibinfo{pages}{60} (\bibinfo{year}{2023}).

\bibitem{politico2021poland}
\bibinfo{author}{Wanat, Z.}
\newblock \bibinfo{title}{{Poland’s vaccine skeptics create a political
  headache}}.
\newblock \bibinfo{howpublished}{Politico.
  https://www.politico.eu/article/poland-vaccine-skeptic-vax-hesitancy-political-trouble-polish-coronavirus-covid-19/}
  (\bibinfo{year}{2021}).

\bibitem{pew2022americans}
\bibinfo{author}{Funk, Y.~C.}, \bibinfo{author}{Tyson, A.},
  \bibinfo{author}{Pasquini, G.} \& \bibinfo{author}{Spencer, A.}
\newblock \bibinfo{title}{{Pew Research. Americans Reflect on Nation’s
  COVID-19 Response}}.
\newblock
  \bibinfo{howpublished}{https://www.pewresearch.org/science/2022/07/07/americans-reflect-on-nations-covid-19-response/}
  (\bibinfo{year}{2022}).

\bibitem{recio2021vaccine}
\bibinfo{author}{Recio-Rom{\'a}n, A.}, \bibinfo{author}{Recio-Men{\'e}ndez, M.}
  \& \bibinfo{author}{Rom{\'a}n-Gonz{\'a}lez, M.~V.}
\newblock \bibinfo{journal}{\bibinfo{title}{Vaccine hesitancy and political
  populism. an invariant cross-european perspective}}.
\newblock {\emph{\JournalTitle{International Journal of Environmental Research
  and Public Health}}} \textbf{\bibinfo{volume}{18}}, \bibinfo{pages}{12953}
  (\bibinfo{year}{2021}).

\bibitem{stoeckel2022politics}
\bibinfo{author}{Stoeckel, F.}, \bibinfo{author}{Carter, C.},
  \bibinfo{author}{Lyons, B.~A.} \& \bibinfo{author}{Reifler, J.}
\newblock \bibinfo{journal}{\bibinfo{title}{The politics of vaccine hesitancy
  in europe}}.
\newblock {\emph{\JournalTitle{European Journal of Public Health}}}
  \textbf{\bibinfo{volume}{32}}, \bibinfo{pages}{636--642}
  (\bibinfo{year}{2022}).

\bibitem{clark2022role}
\bibinfo{author}{Clark, S.~E.}, \bibinfo{author}{Bledsoe, M.~C.} \&
  \bibinfo{author}{Harrison, C.~J.}
\newblock \bibinfo{journal}{\bibinfo{title}{The role of social media in
  promoting vaccine hesitancy}}.
\newblock {\emph{\JournalTitle{Current opinion in pediatrics}}}
  \textbf{\bibinfo{volume}{34}}, \bibinfo{pages}{156--162}
  (\bibinfo{year}{2022}).

\bibitem{kobayashi2022evolution}
\bibinfo{author}{Kobayashi, R.} \emph{et~al.}
\newblock \bibinfo{journal}{\bibinfo{title}{Evolution of public opinion on
  covid-19 vaccination in japan: large-scale twitter data analysis}}.
\newblock {\emph{\JournalTitle{Journal of Medical Internet Research}}}
  \textbf{\bibinfo{volume}{24}}, \bibinfo{pages}{e41928}
  (\bibinfo{year}{2022}).

\bibitem{baron2022politicians}
\bibinfo{author}{Baron, R.~J.} \& \bibinfo{author}{Emanuel, E.~J.}
\newblock \bibinfo{title}{Politicians should not be deciding what constitutes
  good medicine}.
\newblock
  \bibinfo{howpublished}{https://www.statnews.com/2022/03/07/politicians-should-not-be-deciding-what-constitutes-good-medicine/}
  (\bibinfo{year}{2022}).

\bibitem{garimella2018quantifying}
\bibinfo{author}{Garimella, K.}, \bibinfo{author}{Morales, G. D.~F.},
  \bibinfo{author}{Gionis, A.} \& \bibinfo{author}{Mathioudakis, M.}
\newblock \bibinfo{journal}{\bibinfo{title}{Quantifying controversy on social
  media}}.
\newblock {\emph{\JournalTitle{ACM Trans. Social Comput.}}}
  \textbf{\bibinfo{volume}{1}}, \bibinfo{pages}{1--27} (\bibinfo{year}{2018}).

\bibitem{ansari2021worldwide}
\bibinfo{author}{Ansari, M. T.~J.} \& \bibinfo{author}{Khan, N.~A.}
\newblock \bibinfo{journal}{\bibinfo{title}{Worldwide covid-19 vaccines
  sentiment analysis through twitter content.}}
\newblock {\emph{\JournalTitle{Electronic Journal of General Medicine}}}
  \textbf{\bibinfo{volume}{18}} (\bibinfo{year}{2021}).

\bibitem{reshi2022covid}
\bibinfo{author}{Reshi, A.~A.} \emph{et~al.}
\newblock \bibinfo{title}{Covid-19 vaccination-related sentiments analysis: a
  case study using worldwide twitter dataset}.
\newblock In \emph{\bibinfo{booktitle}{Healthcare}}, vol.~\bibinfo{volume}{10},
  \bibinfo{pages}{411} (\bibinfo{organization}{MDPI}, \bibinfo{year}{2022}).

\bibitem{DVN/UKILBE_2022}
\bibinfo{author}{Döring, H.}, \bibinfo{author}{Huber, C.} \&
  \bibinfo{author}{Manow, P.}
\newblock \bibinfo{title}{{ParlGov 2022 Release}},
  \doiprefix\url{10.7910/DVN/UKILBE} (\bibinfo{year}{2022}).

\bibitem{saphore2021white}
\bibinfo{author}{Saphore, S.}
\newblock \bibinfo{title}{White supremacist and far right ideology underpin
  anti-vax movements}.
\newblock
  \bibinfo{howpublished}{https://theconversation.com/white-supremacist-and-far-right-ideology-underpin-anti-vax-movements-172289}
  (\bibinfo{year}{2021}).

\bibitem{broderick2021italy}
\bibinfo{author}{Broderick, R.}
\newblock \bibinfo{title}{Italy’s anti-vaccination movement is militant and
  dangerous}.
\newblock
  \bibinfo{howpublished}{https://foreignpolicy.com/2021/11/13/italy-anti-vaccination-movement-militant-dangerous/}
  (\bibinfo{year}{2021}).

\bibitem{serrano2023far}
\bibinfo{author}{Serrano-Alarc{\'o}n, M.}, \bibinfo{author}{Wang, Y.},
  \bibinfo{author}{Kentikelenis, A.}, \bibinfo{author}{Mckee, M.} \&
  \bibinfo{author}{Stuckler, D.}
\newblock \bibinfo{journal}{\bibinfo{title}{The far-right and anti-vaccine
  attitudes: lessons from spain’s mass covid-19 vaccine roll-out}}.
\newblock {\emph{\JournalTitle{European Journal of Public Health}}}
  \textbf{\bibinfo{volume}{33}}, \bibinfo{pages}{215--221}
  (\bibinfo{year}{2023}).

\bibitem{allington2021health}
\bibinfo{author}{Allington, D.}, \bibinfo{author}{Duffy, B.},
  \bibinfo{author}{Wessely, S.}, \bibinfo{author}{Dhavan, N.} \&
  \bibinfo{author}{Rubin, J.}
\newblock \bibinfo{journal}{\bibinfo{title}{Health-protective behaviour, social
  media usage and conspiracy belief during the covid-19 public health
  emergency}}.
\newblock {\emph{\JournalTitle{Psychological medicine}}}
  \textbf{\bibinfo{volume}{51}}, \bibinfo{pages}{1763--1769}
  (\bibinfo{year}{2021}).

\bibitem{wollebaek2022right}
\bibinfo{author}{Wolleb{\ae}k, D.}, \bibinfo{author}{Fladmoe, A.},
  \bibinfo{author}{Steen-Johnsen, K.} \& \bibinfo{author}{Ihlen, {\O}.}
\newblock \bibinfo{journal}{\bibinfo{title}{Right-wing ideological constraint
  and vaccine refusal: The case of the covid-19 vaccine in norway}}.
\newblock {\emph{\JournalTitle{Scandinavian Political Studies}}}
  \textbf{\bibinfo{volume}{45}}, \bibinfo{pages}{253--278}
  (\bibinfo{year}{2022}).

\bibitem{fischer2023metacognition}
\bibinfo{author}{Fischer, H.}, \bibinfo{author}{Huff, M.},
  \bibinfo{author}{Anders, G.} \& \bibinfo{author}{Said, N.}
\newblock \bibinfo{journal}{\bibinfo{title}{Metacognition, public health
  compliance, and vaccination willingness}}.
\newblock {\emph{\JournalTitle{Proceedings of the National Academy of
  Sciences}}} \textbf{\bibinfo{volume}{120}}, \bibinfo{pages}{e2105425120}
  (\bibinfo{year}{2023}).

\bibitem{bilewicz2022politics}
\bibinfo{author}{Bilewicz, M.} \& \bibinfo{author}{Soral, W.}
\newblock \bibinfo{journal}{\bibinfo{title}{The politics of vaccine hesitancy:
  An ideological dual-process approach}}.
\newblock {\emph{\JournalTitle{Social Psychological and Personality Science}}}
  \textbf{\bibinfo{volume}{13}}, \bibinfo{pages}{1080--1089}
  (\bibinfo{year}{2022}).

\bibitem{klymak2022partisanship}
\bibinfo{author}{Klymak, M.} \& \bibinfo{author}{Vlandas, T.}
\newblock \bibinfo{journal}{\bibinfo{title}{Partisanship and covid-19
  vaccination in the uk}}.
\newblock {\emph{\JournalTitle{Scientific Reports}}}
  \textbf{\bibinfo{volume}{12}}, \bibinfo{pages}{19785} (\bibinfo{year}{2022}).

\bibitem{kouzy2020coronavirus}
\bibinfo{author}{Kouzy, R.} \emph{et~al.}
\newblock \bibinfo{journal}{\bibinfo{title}{Coronavirus goes viral: quantifying
  the covid-19 misinformation epidemic on twitter}}.
\newblock {\emph{\JournalTitle{Cureus}}} \textbf{\bibinfo{volume}{12}}
  (\bibinfo{year}{2020}).

\bibitem{gallotti2020assessing}
\bibinfo{author}{Gallotti, R.}, \bibinfo{author}{Valle, F.},
  \bibinfo{author}{Castaldo, N.}, \bibinfo{author}{Sacco, P.} \&
  \bibinfo{author}{De~Domenico, M.}
\newblock \bibinfo{journal}{\bibinfo{title}{Assessing the risks of
  ‘infodemics’ in response to covid-19 epidemics}}.
\newblock {\emph{\JournalTitle{Nature Human Behaviour}}}
  \textbf{\bibinfo{volume}{4}}, \bibinfo{pages}{1285--1293}
  (\bibinfo{year}{2020}).

\bibitem{ap2023uk}
\bibinfo{author}{{AP News}}.
\newblock \bibinfo{title}{Uk conservatives suspend lawmaker for vaccine
  misinformation}.
\newblock
  \bibinfo{howpublished}{https://apnews.com/article/british-politics-health-united-kingdom-government-f463bd4fdb343a6efb9953ba50b0dfa5}
  (\bibinfo{year}{2023}).

\bibitem{bestvater2022politics}
\bibinfo{author}{Bestvater, S.}, \bibinfo{author}{Shah, S.},
  \bibinfo{author}{Rivero, G.} \& \bibinfo{author}{Smith, A.}
\newblock \bibinfo{title}{Politics on twitter: One-third of tweets from u.s.
  adults are political}.
\newblock \bibinfo{howpublished}{Pew Research.
  https://www.pewresearch.org/politics/2022/06/16/politics-on-twitter-one-third-of-tweets-from-u-s-adults-are-political/}
  (\bibinfo{year}{2022}).

\bibitem{vaccines9060593}
\bibinfo{author}{Jennings, W.} \emph{et~al.}
\newblock \bibinfo{journal}{\bibinfo{title}{Lack of trust, conspiracy beliefs,
  and social media use predict covid-19 vaccine hesitancy}}.
\newblock {\emph{\JournalTitle{Vaccines}}} \textbf{\bibinfo{volume}{9}}
  (\bibinfo{year}{2021}).

\bibitem{sallam2021high}
\bibinfo{author}{Sallam, M.} \emph{et~al.}
\newblock \bibinfo{journal}{\bibinfo{title}{High rates of covid-19 vaccine
  hesitancy and its association with conspiracy beliefs: A study in jordan and
  kuwait among other arab countries}}.
\newblock {\emph{\JournalTitle{Vaccines}}} \textbf{\bibinfo{volume}{9}},
  \bibinfo{pages}{42} (\bibinfo{year}{2021}).

\bibitem{humansubjectsresearch}
\bibinfo{author}{{CITI Program}}.
\newblock \bibinfo{title}{Human subjects research (hsr)}.
\newblock
  \bibinfo{howpublished}{https://about.citiprogram.org/series/human-subjects-research-hsr/,
  how="[Accessed on April 20, 2023]"} (\bibinfo{year}{2023}).

\bibitem{geonames}
\bibinfo{title}{Geonames}.
\newblock \bibinfo{howpublished}{http://www.geonames.org/}.

\bibitem{conover2011political}
\bibinfo{author}{Conover, M.} \emph{et~al.}
\newblock \bibinfo{title}{Political polarization on twitter}.
\newblock In \emph{\bibinfo{booktitle}{Proceedings of the international aaai
  conference on web and social media}}, vol.~\bibinfo{volume}{5},
  \bibinfo{pages}{89--96} (\bibinfo{year}{2011}).

\bibitem{garimella2017effect}
\bibinfo{author}{Garimella, K.}, \bibinfo{author}{De~Francisci~Morales, G.},
  \bibinfo{author}{Gionis, A.} \& \bibinfo{author}{Mathioudakis, M.}
\newblock \bibinfo{title}{The effect of collective attention on controversial
  debates on social media}.
\newblock In \emph{\bibinfo{booktitle}{Proceedings of the 2017 ACM on Web
  Science Conference}}, \bibinfo{pages}{43--52} (\bibinfo{year}{2017}).

\bibitem{casas2017survey}
\bibinfo{author}{Casas-Roma, J.}, \bibinfo{author}{Herrera-Joancomart{\'\i},
  J.} \& \bibinfo{author}{Torra, V.}
\newblock \bibinfo{journal}{\bibinfo{title}{A survey of graph-modification
  techniques for privacy-preserving on networks}}.
\newblock {\emph{\JournalTitle{Artificial Intelligence Review}}}
  \textbf{\bibinfo{volume}{47}}, \bibinfo{pages}{341--366}
  (\bibinfo{year}{2017}).

\bibitem{dall2021nishimori}
\bibinfo{author}{Dall’Amico, L.}, \bibinfo{author}{Couillet, R.} \&
  \bibinfo{author}{Tremblay, N.}
\newblock \bibinfo{journal}{\bibinfo{title}{Nishimori meets bethe: a spectral
  method for node classification in sparse weighted graphs}}.
\newblock {\emph{\JournalTitle{Journal of Statistical Mechanics: Theory and
  Experiment}}} \textbf{\bibinfo{volume}{2021}}, \bibinfo{pages}{093405}
  (\bibinfo{year}{2021}).

\bibitem{dall2021unified}
\bibinfo{author}{Dall'Amico, L.}, \bibinfo{author}{Couillet, R.} \&
  \bibinfo{author}{Tremblay, N.}
\newblock \bibinfo{journal}{\bibinfo{title}{A unified framework for spectral
  clustering in sparse graphs}}.
\newblock {\emph{\JournalTitle{The Journal of Machine Learning Research}}}
  \textbf{\bibinfo{volume}{22}}, \bibinfo{pages}{9859--9914}
  (\bibinfo{year}{2021}).

\bibitem{blondel2008fast}
\bibinfo{author}{Blondel, V.~D.}, \bibinfo{author}{Guillaume, J.-L.},
  \bibinfo{author}{Lambiotte, R.} \& \bibinfo{author}{Lefebvre, E.}
\newblock \bibinfo{journal}{\bibinfo{title}{Fast unfolding of communities in
  large networks}}.
\newblock {\emph{\JournalTitle{Journal of statistical mechanics: theory and
  experiment}}} \textbf{\bibinfo{volume}{2008}}, \bibinfo{pages}{P10008}
  (\bibinfo{year}{2008}).

\bibitem{haman2021politicians}
\bibinfo{author}{Haman, M.} \& \bibinfo{author}{{\v{S}}koln{\'\i}k, M.}
\newblock \bibinfo{journal}{\bibinfo{title}{Politicians on social media. the
  online database of members of national parliaments on twitter}}.
\newblock {\emph{\JournalTitle{Profesional de la informaci{\'o}n}}}
  \textbf{\bibinfo{volume}{30}} (\bibinfo{year}{2021}).

\bibitem{van2020twitter}
\bibinfo{author}{van Vliet, L.}, \bibinfo{author}{T{\"o}rnberg, P.} \&
  \bibinfo{author}{Uitermark, J.}
\newblock \bibinfo{journal}{\bibinfo{title}{The twitter parliamentarian
  database: Analyzing twitter politics across 26 countries}}.
\newblock {\emph{\JournalTitle{PLoS one}}} \textbf{\bibinfo{volume}{15}},
  \bibinfo{pages}{e0237073} (\bibinfo{year}{2020}).

\bibitem{vrandevcic2014wikidata}
\bibinfo{author}{Vrande{\v{c}}i{\'c}, D.} \& \bibinfo{author}{Kr{\"o}tzsch, M.}
\newblock \bibinfo{journal}{\bibinfo{title}{Wikidata: a free collaborative
  knowledgebase}}.
\newblock {\emph{\JournalTitle{Communications of the ACM}}}
  \textbf{\bibinfo{volume}{57}}, \bibinfo{pages}{78--85}
  (\bibinfo{year}{2014}).

\bibitem{jungherr2012pirate}
\bibinfo{author}{Jungherr, A.}, \bibinfo{author}{J{\"u}rgens, P.} \&
  \bibinfo{author}{Schoen, H.}
\newblock \bibinfo{journal}{\bibinfo{title}{Why the pirate party won the german
  election of 2009 or the trouble with predictions: A response to tumasjan, a.,
  sprenger, to, sander, pg, \& welpe, im “predicting elections with twitter:
  What 140 characters reveal about political sentiment”}}.
\newblock {\emph{\JournalTitle{Social science computer review}}}
  \textbf{\bibinfo{volume}{30}}, \bibinfo{pages}{229--234}
  (\bibinfo{year}{2012}).

\bibitem{ozili2023acceptable}
\bibinfo{author}{Ozili, P.~K.}
\newblock \bibinfo{title}{The acceptable r-square in empirical modelling for
  social science research}.
\newblock In \emph{\bibinfo{booktitle}{Social Research Methodology and
  Publishing Results: A Guide to Non-Native English Speakers}},
  \bibinfo{pages}{134--143} (\bibinfo{publisher}{IGI Global},
  \bibinfo{year}{2023}).

\bibitem{jacopo_lenti_2023_7716817}
\bibinfo{author}{Lenti, J.}
\newblock \bibinfo{title}{{Global misinformation spillovers in the online
  vaccination debate before and during COVID-19}},
  \doiprefix\url{10.5281/zenodo.7716817} (\bibinfo{year}{2023}).

\end{thebibliography}

\section*{Acknowledgements}

The authors acknowledge support from the Lagrange Project of the Institute for Scientific Interchange Foundation (ISI Foundation) funded by Fondazione Cassa di Risparmio di Torino (Fondazione CRT). LD further acknowledges support from Fondation Botnar. GP also acknowledges the project "National Center for HPC, Big Data and Quantum Computing", CN00000013 (Bando M42C – Investimento 1.4 – Avviso Centri Nazionali” – D.D. n. 3138 of 16.12.2021, funded with MUR Decree n. 1031 of 17.06.2022).

\section*{Author contributions}



MT, YM, JL, and GP collected the data, GP conducted the experiments, GP, YM, MS, KK, MT, LD wrote the manuscript. All authors participated in the experimental design, interpretation, and review of the manuscript.

\section*{Data availability}
\label{sec:Data_availability}

The original collected data from previous work \textit{Global misinformation spillovers in the online vaccination debate before and during COVID-19}\cite{lenti2022global} are available on Zenodo \cite{jacopo_lenti_2023_7716817}.
Following the Terms of Service of X (formally known as Twitter), only tweet IDs are shared. 
For the collaborations involving the rest of the data, please contact the corresponding author.
For the list of keywords used to collect the data, the annotation of the political users, the dataset of labeled tweets, and a full listing of OLS's coefficients for each party across periods see \url{https://github.com/GiordanoPaoletti/Political-Issue-or-Public-Health}




\section*{Additional information}
The authors declare no competing interests.

\end{document}


\maketitle

\section{Data Volume}
\label{SM:Data_Volume}

\begin{figure}[ht!]
\centering
\includegraphics[width=0.6\linewidth]{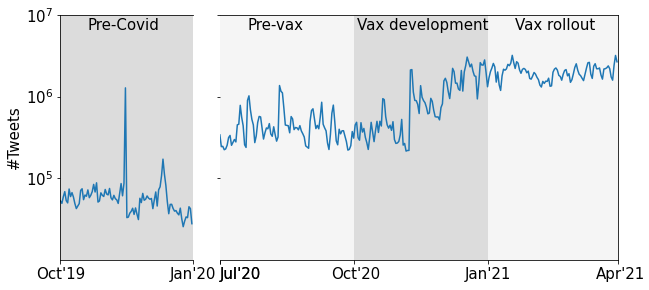}
\caption{Number of tweets over time, four periods identified by background color. 
}
\label{fig:tweet_per_day}
\end{figure}

The volume of the data collected can be seen in the Figure S\ref{fig:tweet_per_day}. We can observe that the volume increases drastically during COVID-19 (note the logarithmic scale on the $y$ axis). 
The missing time frame between periods 1 and 2 encompasses the very start of the COVID-19 worldwide epidemic. 
Unfortunately, it coincides with a technical difficulty in collecting the increased volume of data.
Due to the highly speculative and uncertain nature of that time period, we also exclude it in our analysis of the vaccination debate.

\section{Evaluation of Geo-localization Quality}
\label{SM:Geo_localization}
\begin{table}[ht!]
\centering
\caption{Accuracy of user profile-based geolocation, compared to the GPS coordinates of the post, by official languages of the country and other languages. Instances where fewer than 10 users are available are left blank.}
\label{tab:accuracy_of_geolocalization}
\begin{tabular}{c|cc|cc} \toprule
                                 & \multicolumn{2}{c|}{\textbf{Official lang.}} & \multicolumn{2}{c}{\textbf{Other lang.}} \\ \midrule
\multicolumn{1}{c|}{\textbf{Country}}     & \textbf{\# Users }        & \textbf{Accuracy}         & \textbf{\# Users}       & \textbf{Accuracy}       \\ \midrule
\multicolumn{1}{c|}{\textbf{AT}} & 34               & 0.882353         & 14             & 0.500000       \\
\multicolumn{1}{c|}{\textbf{BE}} & 57               & 0.947368         & 12             & 0.750000       \\
\multicolumn{1}{c|}{\textbf{CH}} & 32               & 0.843750         & 15             & 0.466667       \\
\multicolumn{1}{c|}{\textbf{CZ}} & 7                &               & 9              &             \\
\multicolumn{1}{c|}{\textbf{DE}} & 309              & 0.964401         & 54             & 0.555556       \\
\multicolumn{1}{c|}{\textbf{DK}} & 24               & 1.000000         & 3              &             \\
\multicolumn{1}{c|}{\textbf{ES}} & 525              & 0.927619         & 53             & 0.471698       \\
\multicolumn{1}{c|}{\textbf{FI}} & 27               & 0.962963         & 5              &             \\
\multicolumn{1}{c|}{\textbf{FR}} & 182              & 0.939560         & 62             & 0.209677       \\
\multicolumn{1}{c|}{\textbf{GB}} & 582              & 0.903780         & 93             & 0.483871       \\
\multicolumn{1}{c|}{\textbf{GR}} & 19               & 0.947368         & 8              &             \\
\multicolumn{1}{c|}{\textbf{IE}} & 47               & 0.914894         & 14             & 0.357143       \\
\multicolumn{1}{c|}{\textbf{IT}} & 486              & 0.967078         & 131            & 0.236641       \\
\multicolumn{1}{c|}{\textbf{NL}} & 97               & 0.969072         & 26             & 0.576923       \\
\multicolumn{1}{c|}{\textbf{PL}} & 70               & 1.000000         & 17             & 0.764706       \\
\multicolumn{1}{c|}{\textbf{PT}} & 100              & 0.730000         & 2              &             \\
\multicolumn{1}{c|}{\textbf{SE}} & 79               & 1.000000         & 12             & 0.666667      \\
\bottomrule
\end{tabular}
\end{table}
Our approach of geo-locating users using the Location field of their self-description suffers from several drawbacks, including people lying, writing in non-locations, or having ambiguous or homonymous locations.
Thus, we estimate the accuracy of our geolocation method by comparing it to the information shared by the users that includes the precise geo-coordinates of their location.
We consider such geo-location for users using the official languages of the country (ones that we have selected for this study), or other languages (ones we did not consider).
The list of official languages can be found in Table S\ref{tab:officiallanguages}.
As Table S\ref{tab:accuracy_of_geolocalization} shows, the accuracy for the official languages ranges around 0.95. The countries with lowest accuracy are Portugal (0.73) and Austria (0.88). 
Note that the accuracy for the other languages is much lower, justifying our selection of only those posts which use the official languages of the country.

\section{RT-Networks Sizes}
\label{SM:RT_Networks_Sizes}

Table S\ref{tab:officiallanguages} shows the official languages for each selected country, as well as the number of users in the Giant Weakly Connected Component of the RT Networks for each time period.
We apply a threshold of 300 users to the networks, which excludes 7 networks in the first period, as signified by the lack of numbers in the table.

\begin{table}[h!]
\centering
\caption{Number of nodes in the Giant Weakly Connected Component of the RT-Network of each country/period.}
\label{tab:officiallanguages}
\begin{tabular}{c|c|cccc}
\toprule
\textbf{}        & \textbf{Period}         & \textbf{1} & \textbf{2} & \textbf{3} & \textbf{4} \\
\midrule
\textbf{country} & \textbf{Official lang.} & \multicolumn{4}{c}{\textbf{\# Users in the GCC}}  \\
\midrule

AT & de & &1159&3825&6970 \\
BE & nl, fr, de & &1225&5037&9547 \\
CH & de, fr, it & &837&2727&4774 \\
CZ & cs & &535&2007&3778 \\
DE & de & 5527&11771&32337&58472 \\
DK & da & &468&1496&2625 \\
ES & es & 26634&100284&153274&216996 \\
FI & fi, sv & &905&2114&4242 \\
FR & fr & 17043&34352&92912&148523 \\
GB & en & 24333&98772&269185&392974 \\
GR & el & 790&3331&6461&6657 \\
IE & en & 2850&5498&19209&34053 \\
IT & it & 5514&12819&31979&50169 \\
NL & nl & 1425&5042&13940&20373 \\
PL & pl & 1874&5252&9356&12377 \\
PT & pt & &7482&17682&15598 \\
SE & sv & 654&730&2760&5426 \\
\bottomrule

\end{tabular}
\end{table}

\section{VHE Score Robustness}
\label{SM:VHErobustness}

\begin{figure}[ht!]
\centering
\includegraphics[width=\linewidth]{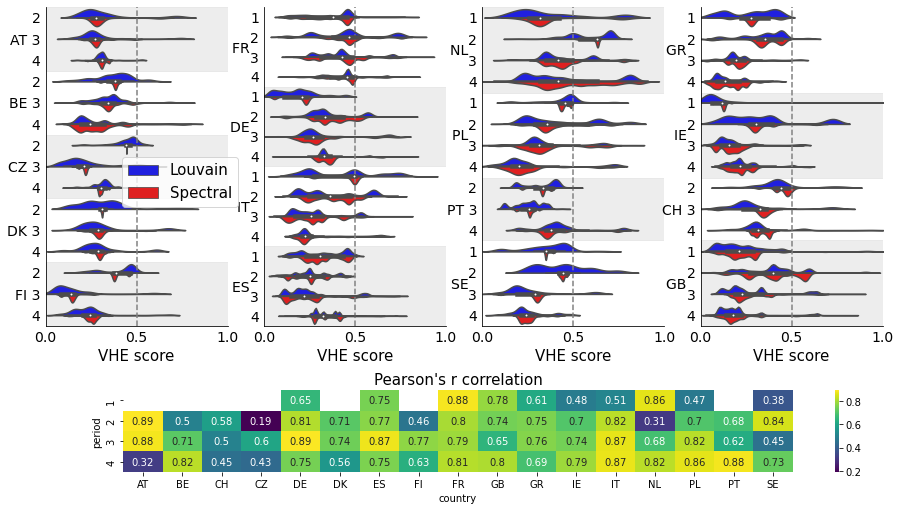}
\caption{Comparison of the VHE scores computed using Louvain and Spectral Clustering as community detection algorithm. White cells in the heatmap indicate country/periods having fewer than 300 users, while all the Pearson's r are significant at $p<10^{-5}$.
}
\label{fig:Louvain_VS_Spectral}
\end{figure}

We check whether the VHE score median and mean correlate with the number of users in the country, and find no such correlation. However, VHE score standard deviation has a positive correlation (0.648) with the number of users. This is expected, because the larger the network, the larger the number of communities that the CD algorithm can find.

We also test the robustness of the VHE score assignment to the choice of community detection algorithm.
We repeat the computation of users' VHE scores using Louvain algorithm, and the manually annotated tweets used in the paper (sampled using Spectral Clustering stratification). 
As can be seen from the Figure S\ref{fig:Louvain_VS_Spectral}, the distributions' shape obtained with Louvain resembles that of Spectral Clustering. 
Notably, there is a strong linear relationship between the two scores, as expressed by Person's $r$ shown in the heatmap (a median of $r$=0.741).
The difference becomes more pronounced for smaller networks.
In general, a higher variance in the scores' distribution is expected, since this method looks for partitions with a high number of communities, without, moreover, giving the possibility of establishing an upper-bound on this  number, like that of $k=15$ which we imposed to Spectral Clustering for interpretation purposes. 
The ability to choose the number of communities also allowed us to perform a stratified sampling of the tweets for annotation, which would have been difficult with Louvain's output.
Given the results above, since to answer our RQs we use linear regression and correlations, the results would likely be similar if Louvain algorithm was used instead.

\section{Results with Louvain-based VHE score}
\label{SM:Louvain_Results}

\begin{figure}[h]
\centering
   \centering  \includegraphics[width=0.45\linewidth]{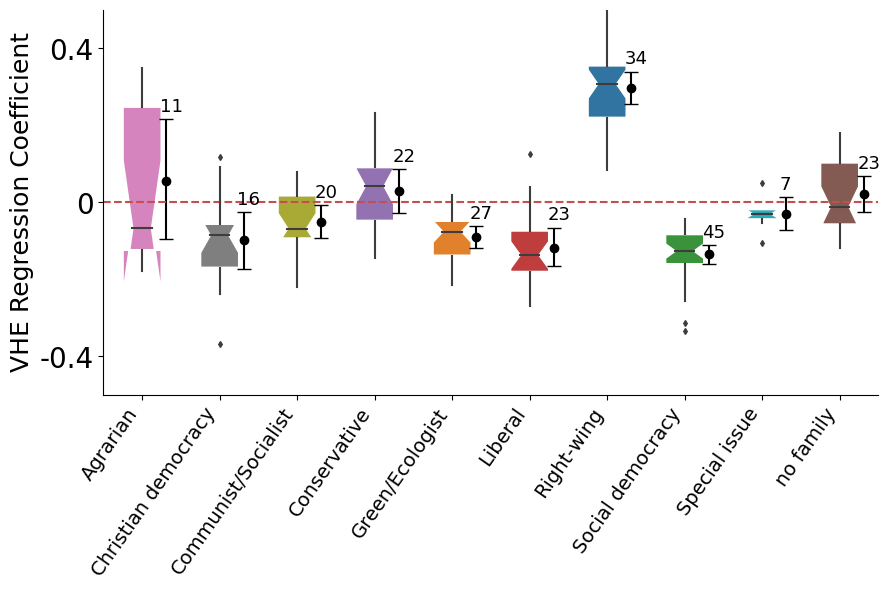}
    \caption{Distribution (boxplots) of OLS coefficients modeling users' VHE score by their interest in parties, grouped by families using ParlGov. Accompanying points and whiskers indicate a 99\% bootstrapped confidence interval. Numbers indicate how many parties are in each group.}
\label{fig:party_family_Louvain}
\end{figure}

To make sure the results are robust to the selection of the clustering algorithm, we computed the main results for Research Questions 1 \& 2, as can be seen in Figures S\ref{fig:party_family_Louvain}-S\ref{fig:Interest_Focus_Heatmap_Louvain}. 
Overall, the main findings do not change, compared to those shown in the main manuscript which use Spectral Clustering.
Those most affected by the selection of the algorithm are the least significant results in Figure S\ref{fig:Interest_Focus_Heatmap_Louvain}, however they do not change the main result of the RQ, mainly that there are few countries having a strong bias in terms of politicization and exposure to vaccine-hesitant content. 

\begin{figure*}[t]
\centering
  \includegraphics[width=0.9\linewidth]{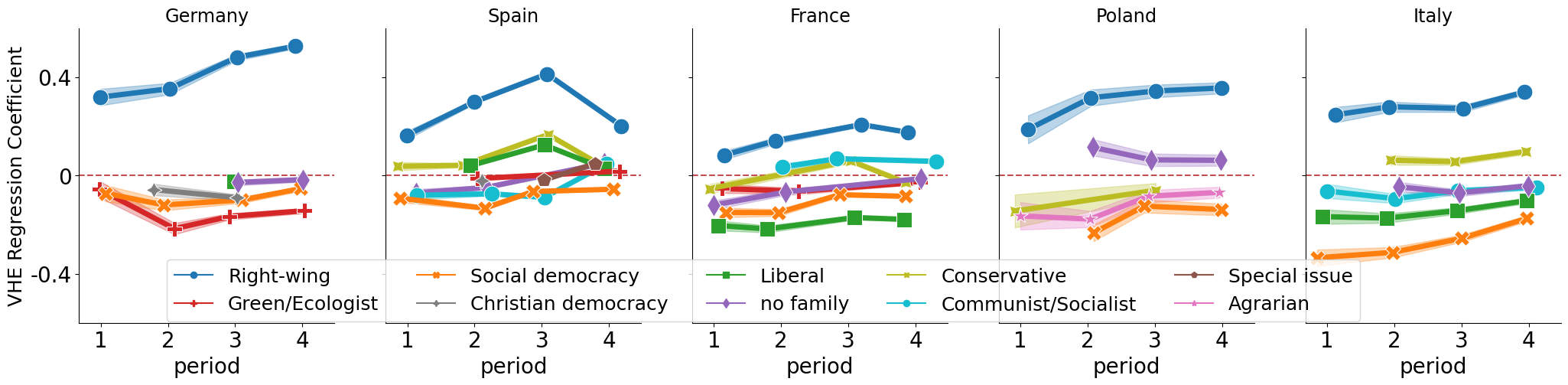}
\caption{Significant OLS coefficients (at $p<0.01$ with Bonferroni correction) for user interest in parties. grouped in families using ParlGoV, and their 99\% confidence intervals. Showing countries having sufficient model fit over 4 time periods. }
\label{fig:stability_trends_Louvain}
\end{figure*}

\begin{figure*}[t]
  \centering
  \includegraphics[width=0.85\linewidth]{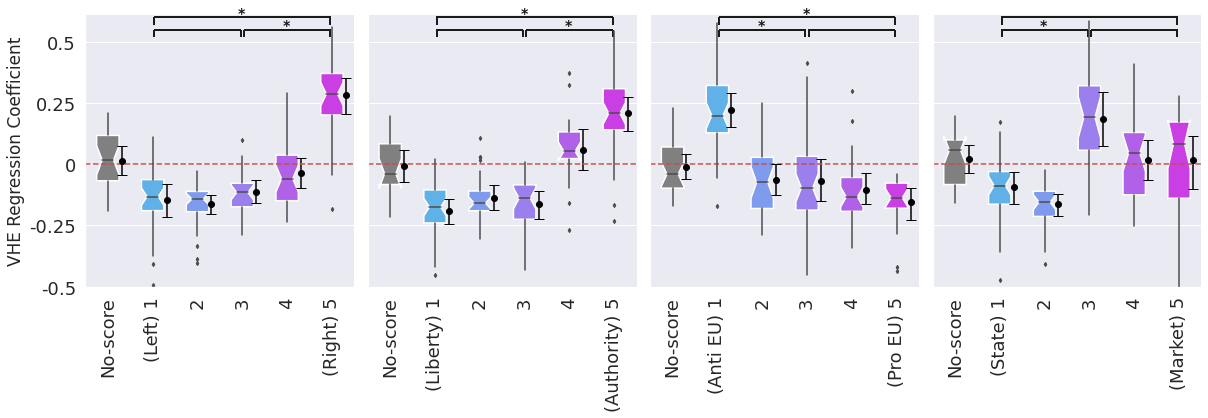}
\caption{Distribution (boxplots) of OLS coefficients modeling users' VHE score by their interest in parties having one of four dimensions defined by ParlGov, grouped in quintiles. Accompanying points and whiskers indicate a 99\% bootstrapped confidence interval. Horizontal brackets on top indicate the comparisons among quintiles $1$, $3$ and $5$, * signifies whether one distribution is statistically greater than the other (one-sided Mann-Whitney U test at $p<0.01$ with Bonferroni correction).}
\label{fig:fourdimensions_Louvain}
\end{figure*}

\begin{figure} [t]
\centering
\includegraphics[width=0.98\linewidth]{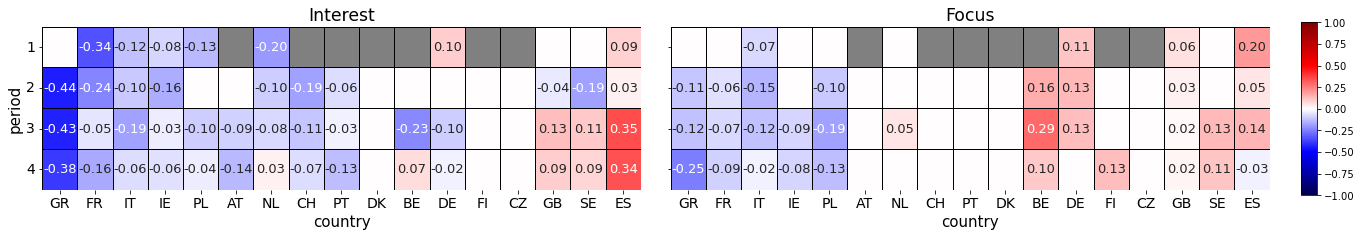}
\caption{Spearman correlation between political interest and VHE score (left) and political focus and VHE score (right) (see RQ2 Methods). Grey cells indicate country/periods having fewer than 300 users. White cells indicate a non-significant correlation.
}
\label{fig:Interest_Focus_Heatmap_Louvain}
\end{figure}

\rev{\section{VHE score validation}}
\label{SM:VHEvalidation}

\begin{figure}[!t]
    \centering
    \includegraphics[width=0.8\linewidth]{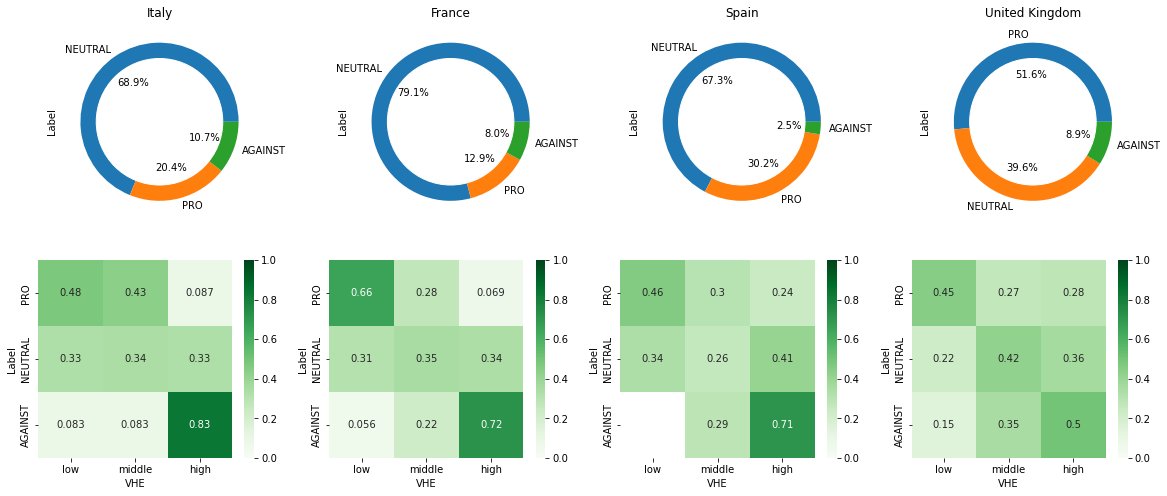}
    \caption{\rev{Summary of annotation results validating the VHE score. Ring plots show the percentage distribution of tweets per label, while heatmaps illustrate how these percentages are distributed among the terciles of VHE score distribution.}}
    \label{fig:VHE_validation}
\end{figure}

\rev{To validate the computation of Vaccine Hesitancy Endorsement score per user, we perform a manual annotation exercise.
For four of the largest countries in our dataset, Italy, France, Spain, and the United Kingdom, we stratify the VHE score into terciles. 
From each tericle we sample 15 users that have posted (or retweeted) at least 5 tweets, and annotate a sample of 5 tweets from each. 
We annotated a total of 900 tweets, and computed the difference between anti- and pro-vax tweets for each user and correlated it to their VHE score.
As for one country, we found only 3 anti-vax tweets, we sampled further 5 users from the first and fifth quintiles. 
The summary of the annotation and the Spearman correlation with the VHE scores can be seen in Figure S\ref{fig:VHE_validation}. 
The correlations are 0.56 ($p<0.001$) for Italy, 0.55 ($p<0.001$) for France, 0.38 ($p<0.01$) for Spain and 0.38 ($p<0.01$) for the United Kingdom.
As can be seen from the figure, out of all anti-vax tweets, 50-83\% were for users having VHE score in the highest tercile. 
We conclude that the proposed approach is a good proxy for the stated endorsement of the no-vax stance.
}

\rev{\section{Details of Linear Regression}
\label{SM:RegressionDetails}

In this section, we offer a concise overview of the implementation specifics for the Ordinary Least Squares (OLS) fit. This encompasses the variables under consideration, starting from the presentation of the fitting equation, which can be decomposed as follows:}

\rev{
$$
VHE^{(u)} \sim \beta_0 + \sum_{i=1}^K\beta_i X_i^{(u)} + \sum_{j=K+1}^N\beta_j Y_j^{(u)}
$$
}
\rev{
Where }
\begin{enumerate}
    \item \rev{$\beta$ represents the weights to be fitted,}
    \item \rev{ $X^{(u)} = \{X_i^{(u)}\}_{i=1}^{k}$is the set of confounding variables, and this feature set remains constant across the various fits. The confounding variables include: the number of followers and followees, daily posting rate, weighted in-degree and weighted out-degree (number of retweets they had, and number of retweets they made in the vaccine debate, respectively), and the proportion of followed users who are politicians (political interest, defined above).}
    \item \rev{$Y^{(u)} = \{Y_j^{(u)}\}_{j=K+1}^{J}$ is the set of features describing the political interest of the user $u$. Initially, for each country, this set is unique and comprises the fraction of politicians followed for each party in the dataset. The complete list of parties for each country is available at the following link: \url{https://github.com/GiordanoPaoletti/Political-Issue-or-Public-Health}.
    For example, when modeling the VHE scores of users in Germany in a particular period, parties considered are \textit{Alliance 90 / Greens}, \textit{Alternative for Germany}, \textit{Christian Democratic Union / Christian Social Union}, \textit{Free Democratic Party}, \textit{Other}, \textit{PDS - The Left}, \textit{Social Democratic Party of Germany}, \textit{independent politician}}.
\end{enumerate}


\rev{For Figures 2 and 3, we group the political parties by party family, as indicated by ParlGov. The resulting list of fitting variables is: number\_of\_followers, daily\_posting\_rate, weighted\_in\_degree, political\_interest, \textit{Conservative}, \textit{Social democracy}, \textit{Green/Ecologist, Liberal}, 
\textit{Christian democracy},  \textit{Communist/Socialist},  \textit{Agrarian}, \textit{no family}, \textit{Right-wing}, \textit{Special issue}.
}

\rev{For Figure 4, we group the political parties by quintiles in each of the four dimensions specified by ParlGov. This would result in four regressions, one for each dimension: left-right, liberty-authority, anti - pro EU, and state-market. So, keeping the confounding variables X unchanged, we define a set of 6 features Y (i.e. one for each quintile and the sixth to group parties with an unknown score).
}

\begin{figure}[!t]
    \centering
    \includegraphics[width=0.7\linewidth]{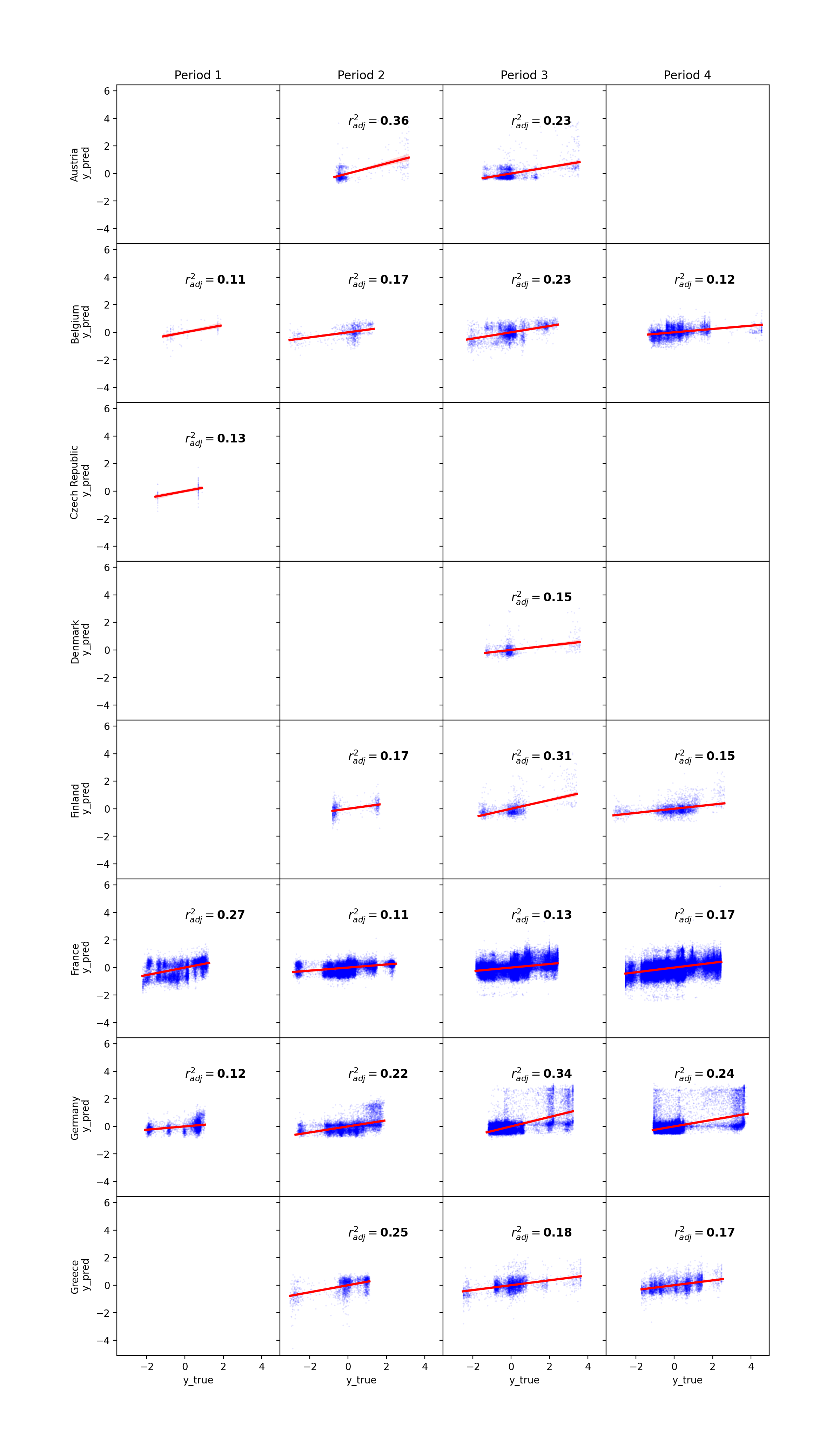}
    \caption{Regression performance modeling VHE score by parties (part 1). The x-axes represent the real target $y_{true}$ (user's VHE score) and the y-axis the predictions from OLS models ($y_{pred}$). }
    \label{fig:VHE_scatter_a}
\end{figure}

\rev{Figures S\ref{fig:VHE_scatter_a} -  S\ref{fig:VHE_scatter_state_market_b} show the scatterplots representing the performance of the regression models described above. In each, the x axis is the true VHE score of the user and the y axis is the model's prediction. Shown are only the regressions in which Adj $R^2 \geq 0.1$ (same threshold is used in the paper for analysis). Note that the score extends beyond the 0 to 1 range because we standardize the target variable to make a consistent comparison among OLS weights across countries and periods.}

\begin{figure}[!t]
    \centering
    \includegraphics[width=0.65\linewidth]{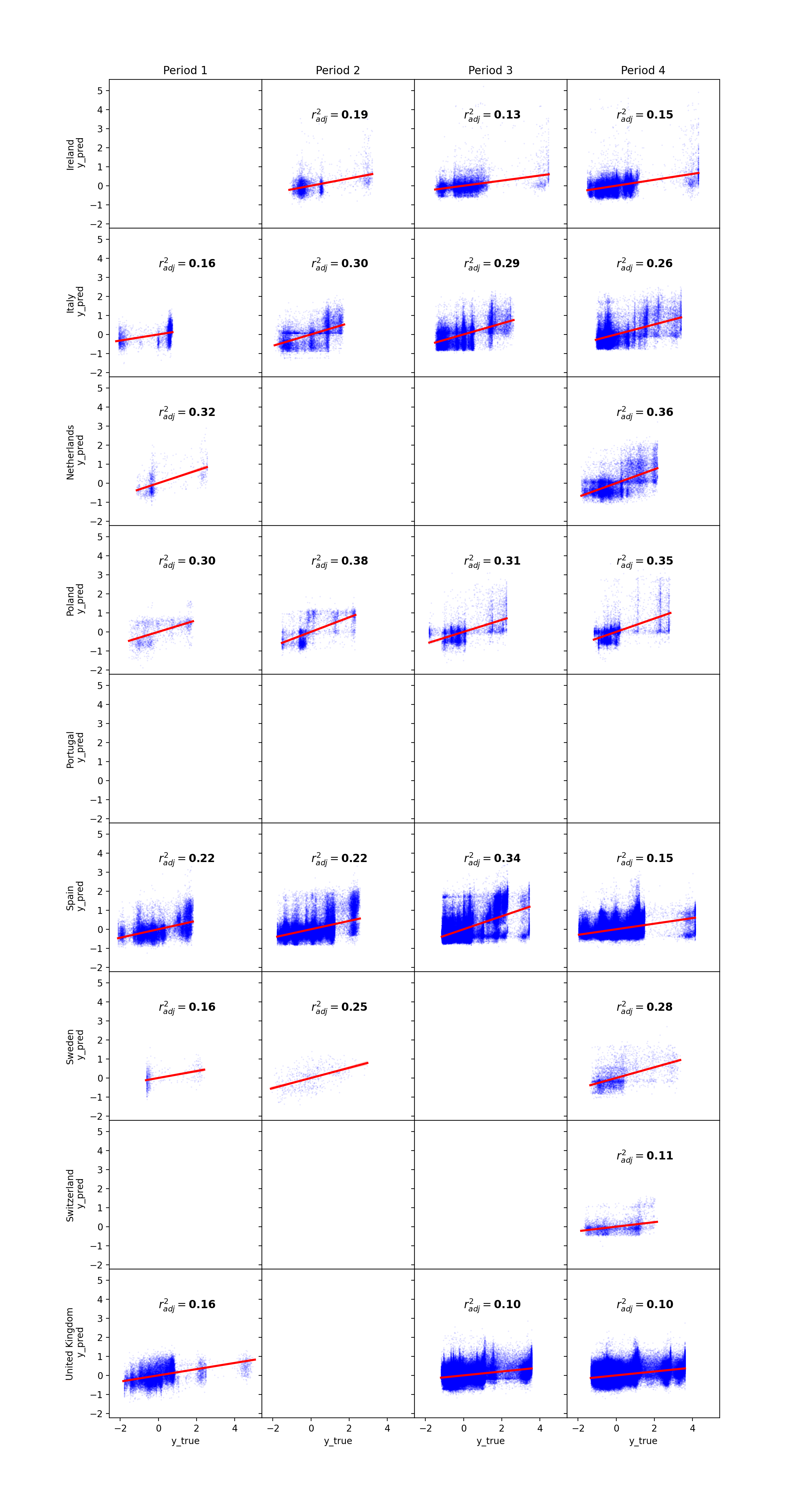}
    \caption{Regression performance modeling VHE score by parties (part 2). The x-axes represent the real target $y_{true}$ (user's VHE score) and the y-axis the predictions from OLS models ($y_{pred}$)}
    \label{fig:VHE_scatter_b}
\end{figure}

\begin{figure}[!t]
    \centering
    \includegraphics[width=0.7\linewidth]{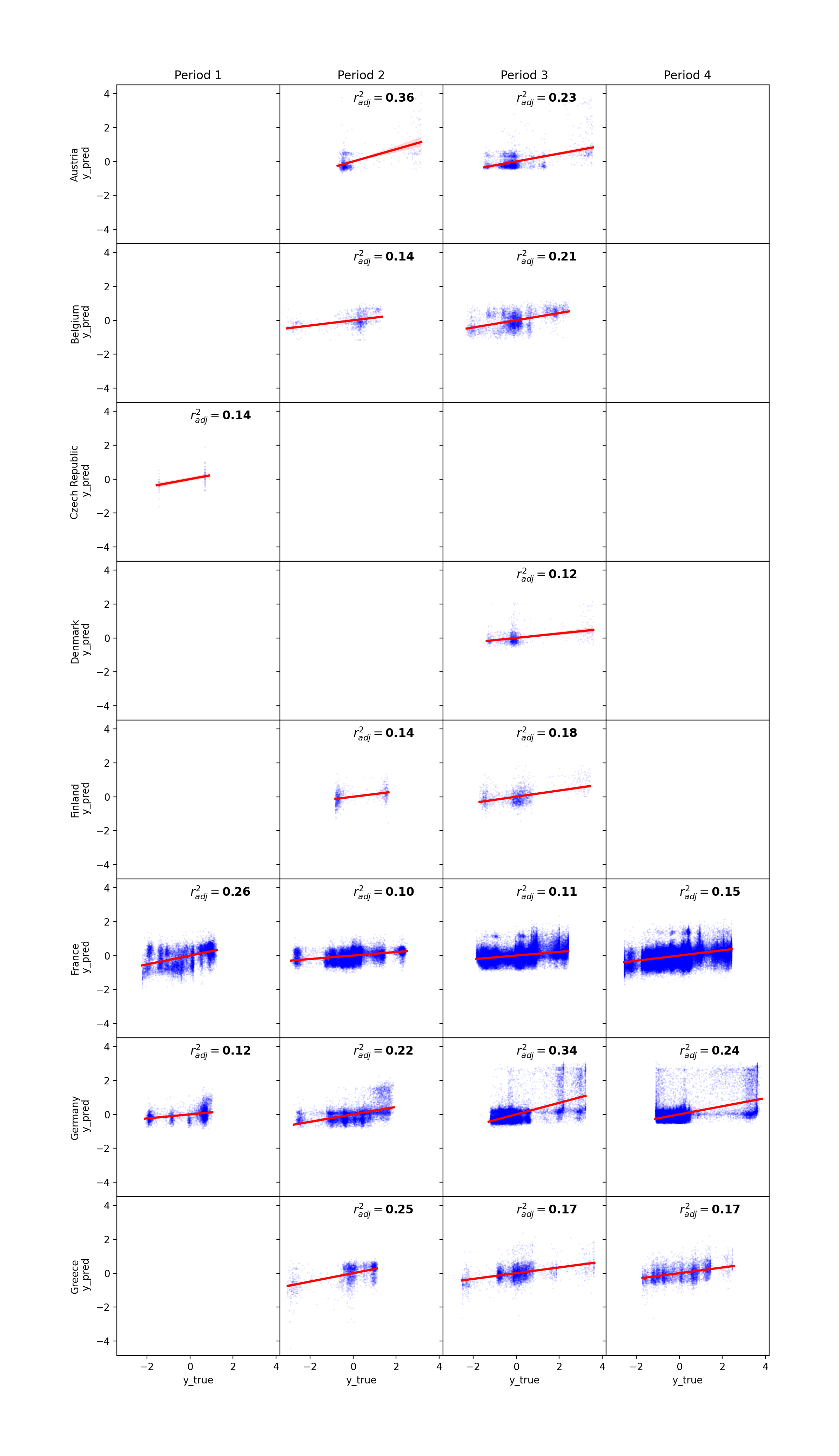}
    \caption{Regression performance modeling VHE score by parties grouped by families using ParlGov,  (part 1). The x-axes represent the real target $y_{true}$ (user's VHE score) and the y-axis the predictions from OLS models ($y_{pred}$). }
    \label{fig:VHE_scatter_family_a}
\end{figure}
\begin{figure}[!t]
    \centering
    \includegraphics[width=0.65\linewidth]{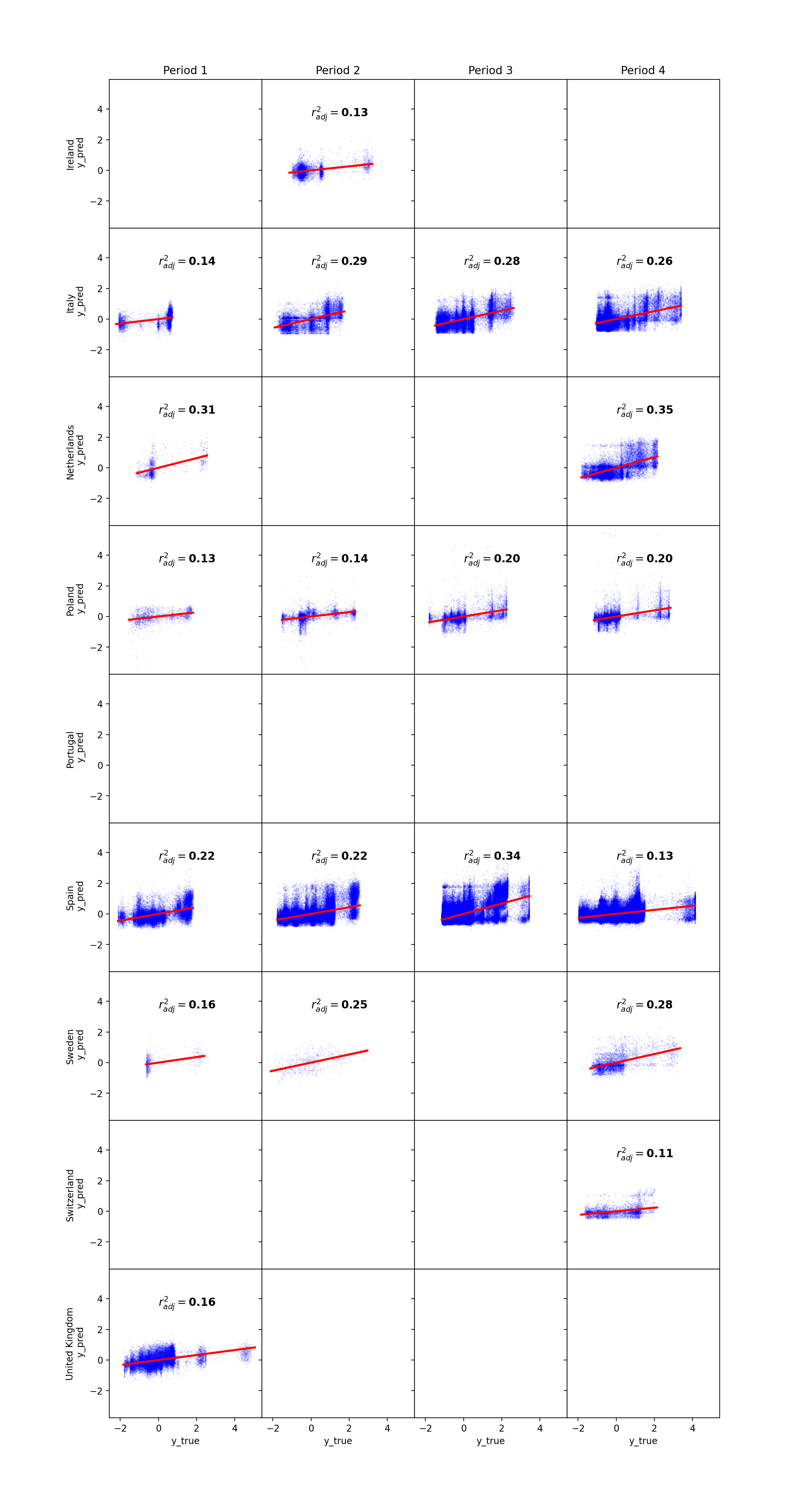}
    \caption{Regression performance modeling VHE score by parties grouped by families using ParlGov,  (part 2). The x-axes represent the real target $y_{true}$ (user's VHE score) and the y-axis the predictions from OLS models ($y_{pred}$). }
    \label{fig:VHE_scatter_family_b}
\end{figure}


\begin{figure}[!t]
    \centering
    \includegraphics[width=0.7\linewidth]{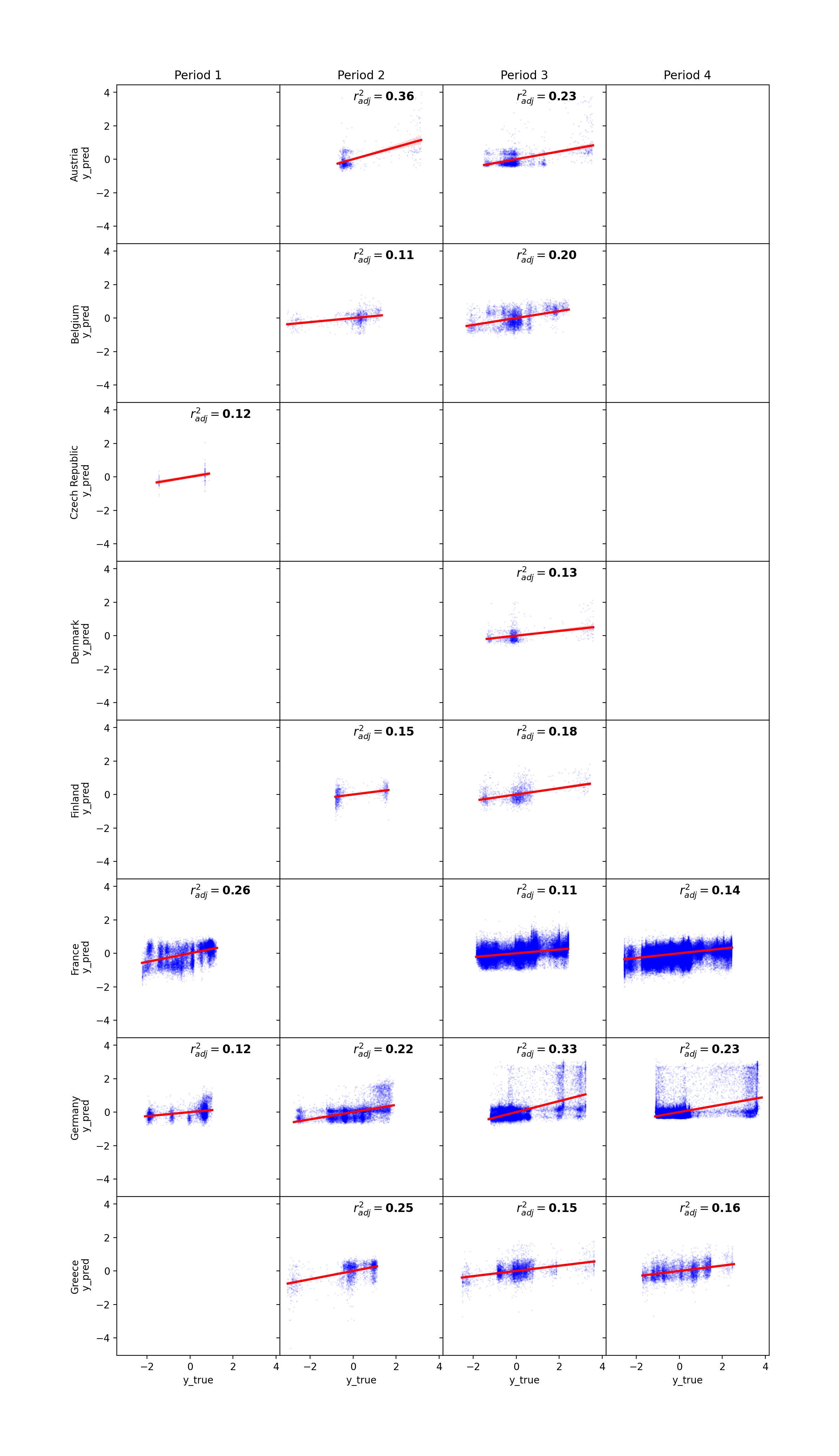}
    \caption{Regression performance modeling VHE score by parties grouped by quintiles of the left-right dimension specified by ParlGov  (part 1). The x-axes represent the real target $y_{true}$ (user's VHE score) and the y-axis the predictions from OLS models ($y_{pred}$). }
    \label{fig:VHE_scatter_left_right_a}
\end{figure}
\begin{figure}[!t]
    \centering
    \includegraphics[width=0.65\linewidth]{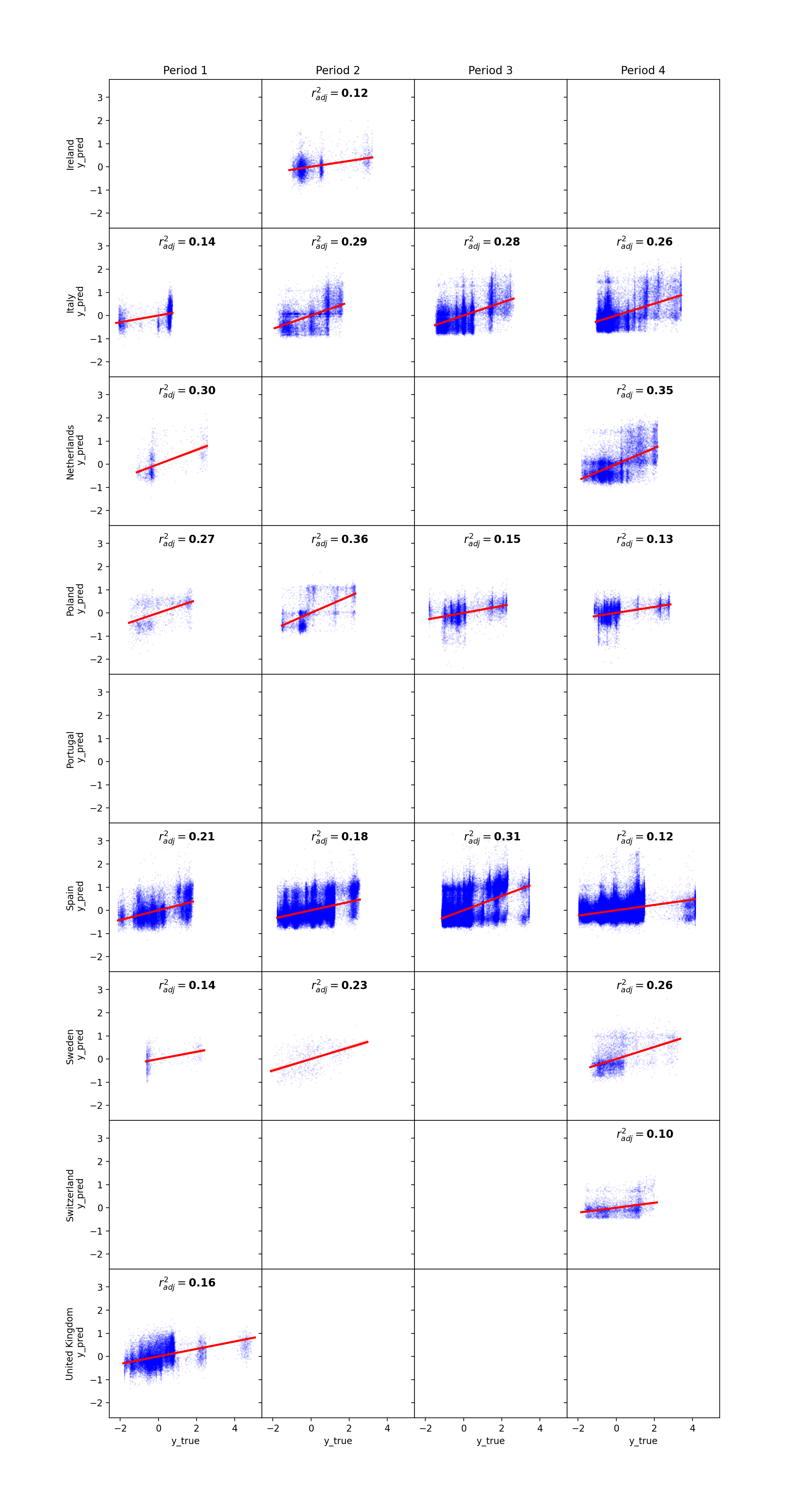}
    \caption{Regression performance modeling VHE score by parties grouped by quintiles of the left-right dimension specified by ParlGov (part 2). The x-axes represent the real target $y_{true}$ (user's VHE score) and the y-axis the predictions from OLS models ($y_{pred}$). }
    \label{fig:VHE_scatter_left_right_b}
\end{figure}


\begin{figure}[!t]
    \centering
    \includegraphics[width=0.7\linewidth]{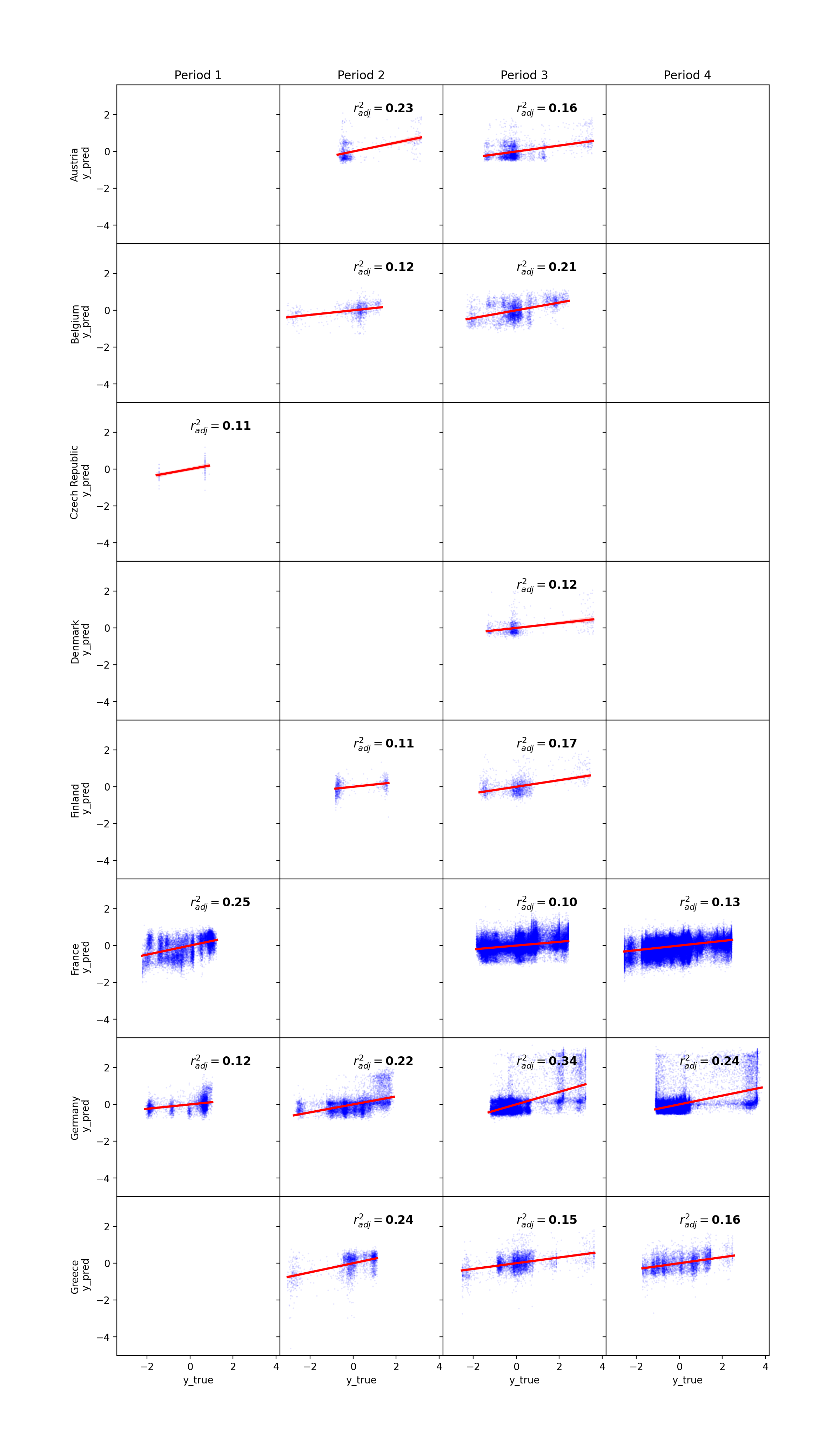}
    \caption{Regression performance modeling VHE score by parties grouped by quintiles of the liberty-authority dimension specified by ParlGov  (part 1). The x-axes represent the real target $y_{true}$ (user's VHE score) and the y-axis the predictions from OLS models ($y_{pred}$). }
    \label{fig:VHE_scatter_liberty_authority_a}
\end{figure}
\begin{figure}[!t]
    \centering
    \includegraphics[width=0.65\linewidth]{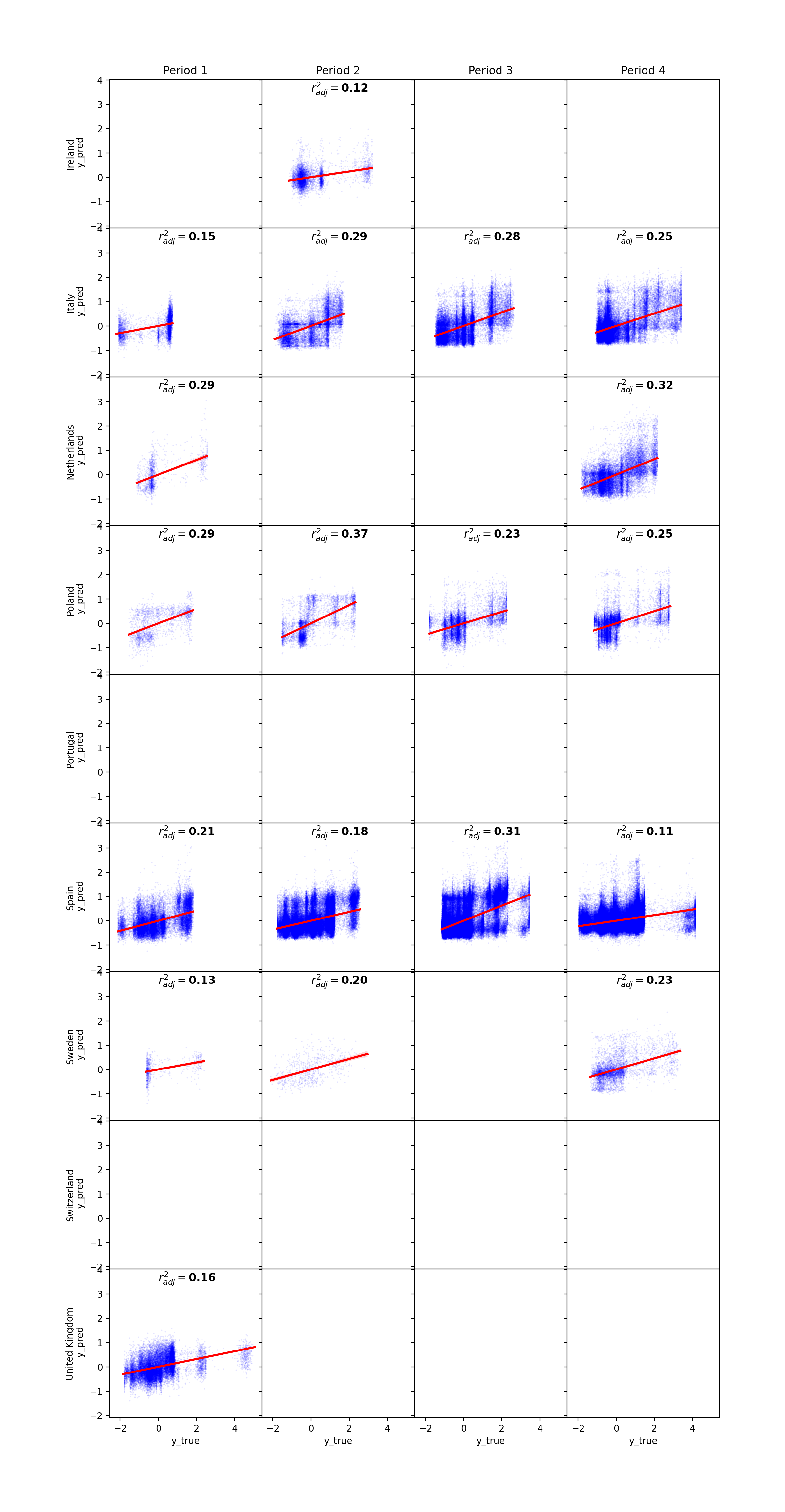}
    \caption{Regression performance modeling VHE score by parties grouped by quintiles of the liberty-authority dimension specified by ParlGov (part 2). The x-axes represent the real target $y_{true}$ (user's VHE score) and the y-axis the predictions from OLS models ($y_{pred}$). }
    \label{fig:VHE_scatter_liberty-authority_b}
\end{figure}


\begin{figure}[!t]
    \centering
    \includegraphics[width=0.7\linewidth]{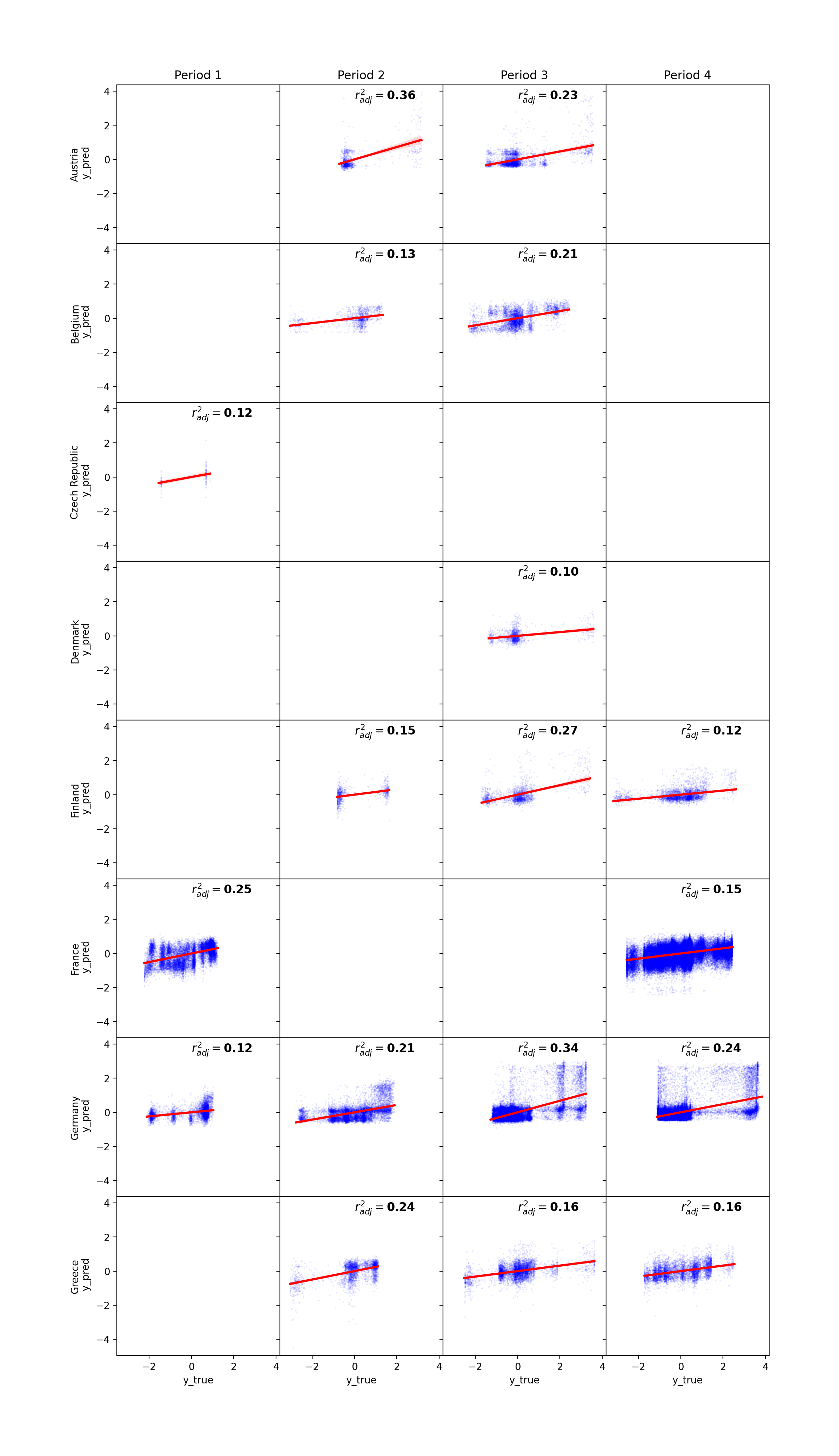}
    \caption{Regression performance modeling VHE score by parties grouped by quintiles of the anti EU - pro EU dimension specified by ParlGov  (part 1). The x-axes represent the real target $y_{true}$ (user's VHE score) and the y-axis the predictions from OLS models ($y_{pred}$). }
    \label{fig:VHE_scatter_anti_pro_eu_a}
\end{figure}
\begin{figure}[!t]
    \centering
    \includegraphics[width=0.65\linewidth]{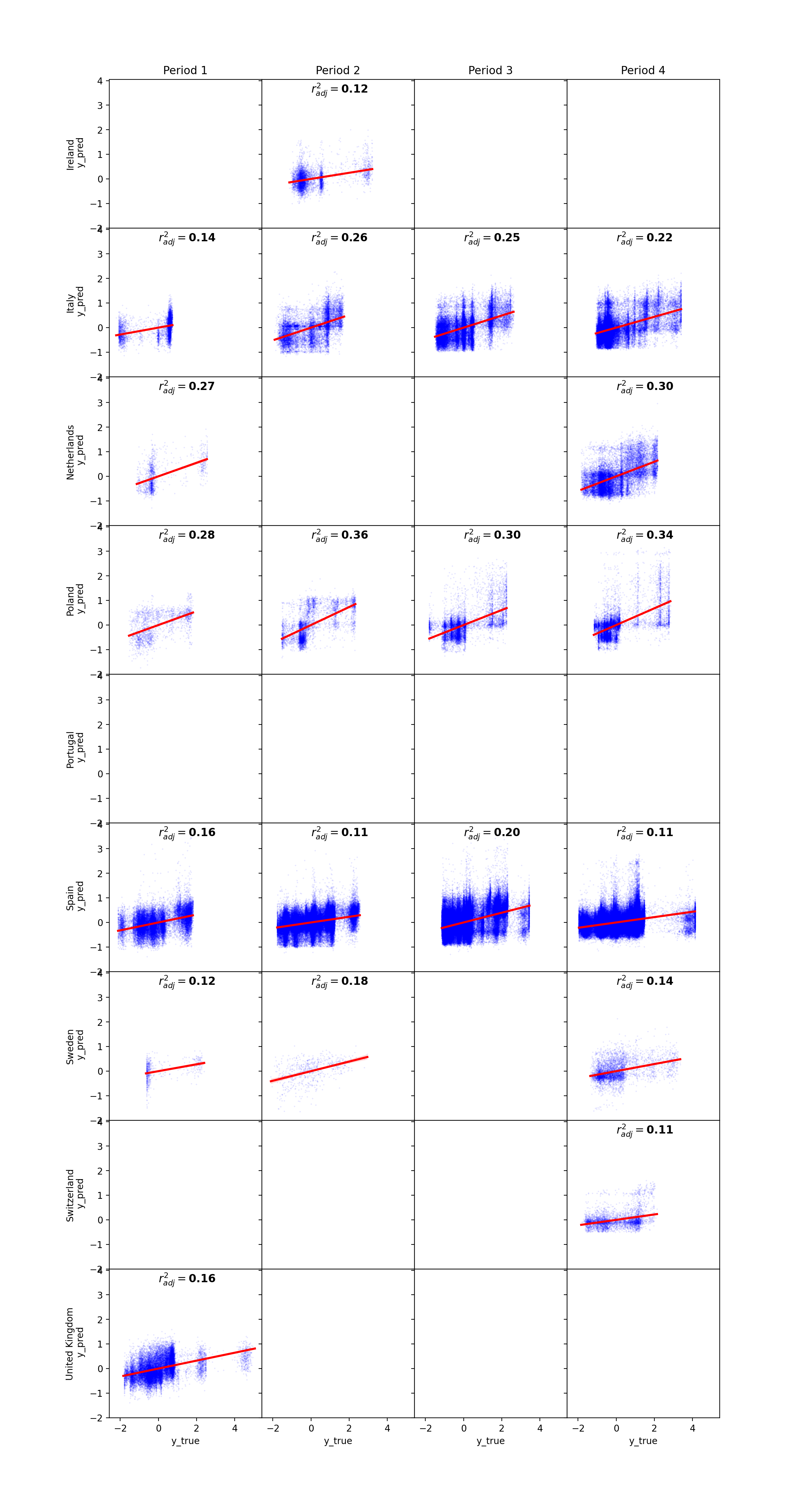}
    \caption{Regression performance modeling VHE score by parties grouped by quintiles of the anti EU - pro EU dimension specified by ParlGov (part 2). The x-axes represent the real target $y_{true}$ (user's VHE score) and the y-axis the predictions from OLS models ($y_{pred}$). }
    \label{fig:VHE_scatter_anti_pro_eu_b}
\end{figure}


\begin{figure}[!t]
    \centering
    \includegraphics[width=0.7\linewidth]{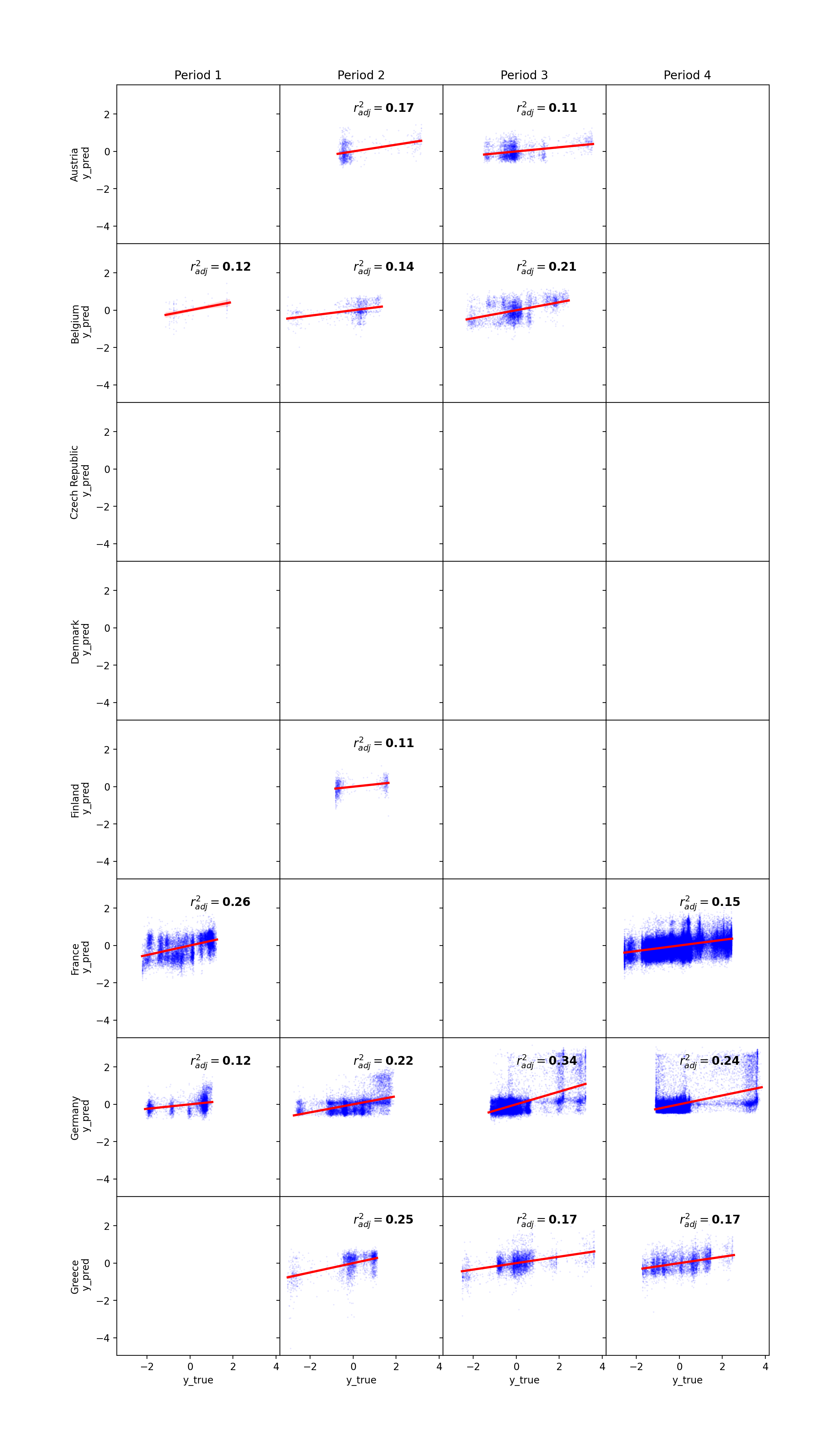}
    \caption{Regression performance modeling VHE score by parties grouped by quintiles of the state-market dimension specified by ParlGov  (part 1). The x-axes represent the real target $y_{true}$ (user's VHE score) and the y-axis the predictions from OLS models ($y_{pred}$). }
    \label{fig:VHE_scatter_state_market_eu_a}
\end{figure}
\begin{figure}[!t]
    \centering
    \includegraphics[width=0.65\linewidth]{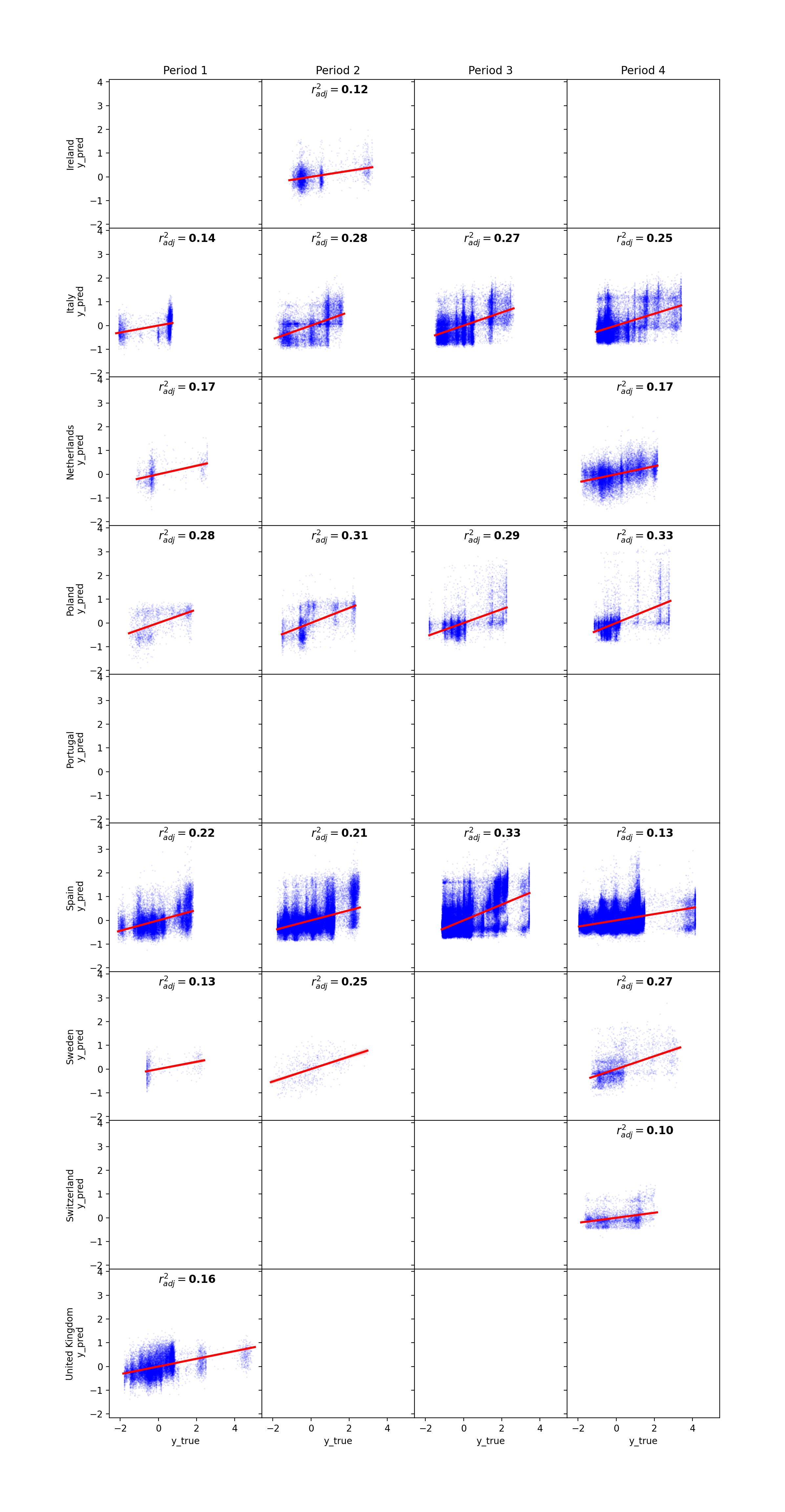}
    \caption{Regression performance modeling VHE score by parties grouped by quintiles of the state-market dimension specified by ParlGov (part 2). The x-axes represent the real target $y_{true}$ (user's VHE score) and the y-axis the predictions from OLS models ($y_{pred}$). }
    \label{fig:VHE_scatter_state_market_b}
\end{figure}